\documentclass[]{JHEP3}

\usepackage{amsfonts}
\usepackage{amsmath,amssymb}
\usepackage{graphicx}

\newcommand{\lb}[0] { \left( }
\newcommand{\rb}[0] { \right) }
\newcommand{\beqs} { \begin{eqnarray} }
\newcommand{\eeqs} { \end{eqnarray} }
\newcommand{\bsub} { \begin{subequations} }
\newcommand{\esub} { \end{subequations} }
\newcommand{\nn} {\nonumber}
\newcommand{\ep}[0] { \epsilon }

\newcommand{\EE}[2] {#1 \times 10^{#2}}
\newcommand{\degree}{\ensuremath{^\circ}}

\newcommand{\Hnull}{\mathcal{H}_{\rm null}}
\newcommand{\Halt}{\mathcal{H}_{\rm alt}}

\newcommand{\Dnull}{\mathcal{D}_{\rm null}}
\newcommand{\Dalt}{\mathcal{D}_{\rm alt}}
\newcommand{\model}{matter tracer model}
\newcommand{\modelsp}{matter tracer model }

\title{Testing large-scale (an)isotropy of ultra-high energy cosmic rays}

\author{Hylke B. J. Koers \\Service de Physique Th\'eorique, Universit\'e Libre de Bruxelles (U.L.B.),
CP225, Bld. du Triomphe, B-1050 Bruxelles, Belgium\\E-mail: \email{hkoers@ulb.ac.be}}
\author{Peter Tinyakov \\
Service de Physique Th\'eorique, Universit\'e Libre de Bruxelles (U.L.B.),
CP225, Bld. du Triomphe, B-1050 Bruxelles, Belgium\\
Institute for Nuclear Research,
60th October Anniversary Prospect 7a, 117312, Moscow, Russia
\\E-mail: \email{petr.tiniakov@ulb.ac.be}}

\abstract{
We present a simple yet powerful method to test
models of cosmic-ray (CR) origin using the distribution of
CR arrival directions. The method is 
statistically unambiguous in the sense that it is
binless and does not invoke scanning over unknown parameters,
and general 
in the sense that it can be applied to any model that predicts a continuous
distribution of CRs over the sky.
We show that it provides a powerful
discrimination between an isotropic distribution
and predictions from the ``matter tracer'' model,
a benchmark model
that assumes small CR deflections and a continuous distribution of sources
tracing the distribution of matter in the Universe.
Our method is competitive or superior in statistical power to existing
methods, and is especially
sensitive in the case of relatively few high-energy
events.  Applying the method to the present data we find that
neither an isotropic distribution nor the \modelsp can be excluded.
Based on estimates of its statistical power, we expect that the
proposed test will lead to meaningful constraints on models of
CR origin with the data that will be accumulated within the
next few years by the Pierre Auger Observatory and the Telescope Array.
}


\preprint{ULB-TH/08-38}

\begin{document}

\section{Introduction}
\label{sec:intro}
The recent observation \cite{Abbasi:2007sv,Abraham:2008ru} of
a strong suppression in the energy spectrum of ultra-high energy
cosmic rays (UHECRs) provides compelling evidence that
the bulk of UHECRs come from relatively
close (within $\sim$$100$ Mpc) extragalactic sources. 
It seems inevitable that the distribution of cosmic-ray (CR) sources
traces the distribution of matter on these length scales.
One should
therefore expect anisotropy in the UHECR flux due to the
non-uniform distribution of matter in the nearby Universe.
Although other experiments have not detected deviations from isotropy at the highest energies,
the Pierre Auger Observatory (PAO) has recently reported 
anisotropy in the flux of UHECRs  above 57 EeV
at a confidence level of more than 99\% \cite{Cronin:2007zz, Abraham:2007si}.

Exactly how CR anisotropy manifests itself depends
crucially on two factors:
the deflections of CRs in magnetic
fields and the density of sources.  The deflections of UHECRs in 
Galactic and extragalactic magnetic fields may vary from a few to a
few tens of degrees, the uncertainty being due to both unknown
parameters of the magnetic fields and the uncertain chemical composition
of CRs at the highest energies. The extragalactic magnetic
fields are usually assumed to be small ($\lesssim 10^{-9}$~G) in voids,
and larger in clusters of galaxies and filaments. Simulations show
that deflections in the extragalactic magnetic fields may be small for
protons for most directions on the sky \cite{Dolag:2004kp} (see,
however, Ref. \cite{Sigl:2004yk}). For iron they may reach several tens of
degrees. The deflections in the Galactic magnetic field may also vary
from a few to a few tens of degrees, depending on the chemical composition
of UHECRs. However, even the largest deflections are unlikely to
completely wash out the anisotropy created by the non-uniform source
distribution.

Although the CR source density affects the anisotropy signature,
deviations from isotropy are expected for any density:
If the number of sources of observable CRs is large, 
anisotropies in the CR flux arise from the fact that
the distribution of sources is inhomogeneous because it
traces that of visible matter.
If, on the other hand, there are few sources, they will produce anisotropy just
because of that. Anisotropy is therefore expected for any source density.

Different assumptions on deflections and source densities lead to
different predictions for the UHECR anisotropy.  The case of small
deflections is of particular interest because in this case the
uncertainties related to magnetic fields are not present.
Under the
assumption of small deflections, the existing data imply that the
number of observable sources above $\sim$$60$ EeV is large, namely several hundred or more
\cite{Abraham:2007si, Dubovsky:2000gv,Kachelriess:2004pc}. This
situation may be approximated by a continuous distribution of sources
following the matter distribution in the Universe. 
This model, which we refer to as the ``matter tracer'' model,
predicts that local matter overdensities such as 
nearby galaxy superclusters should be visible in UHECRs.
The matter tracer model may serve as a benchmark in anisotropy studies, as
its main ingredients
-- small deflections and a large number of observable sources --
represent a limiting case of many realistic situations.
Apart from details on particle acceleration in CR sources, the model has no free parameters: 
for a given injection spectrum the
distribution of UHECR over the sky is predicted uniquely. By comparing
this prediction to observations, one may test to which extent the
underlying assumptions --- small deflections in the magnetic fields,
in the first place --- are correct.

The prediction of a continuous distribution of CRs over the sky
calls for special statistical tests.
In this paper we propose a
new statistical test suitable in this situation.
Several other methods have been discussed in the literature
\cite{Sigl:2004yk, Kashti:2008bw, Waxman:1996hp, Cuoco:2005yd, Kachelriess:2005uf, Kalashev:2007ph, Sommers:2000us, Singh:2003xr, Smialkowski:2002by, Evans:2001rv, Harari:2008zp, Takami:2008ri}. Our
method is competitive or superior in statistical power to the existing
methods (see Sect. \ref{sec:powercomparison} for a comparison in power
between different methods).
It is especially sensitive in the case of a few high-energy events,
which is a regime of physical interest.

\FIGURE{

\includegraphics[height=4.5cm]{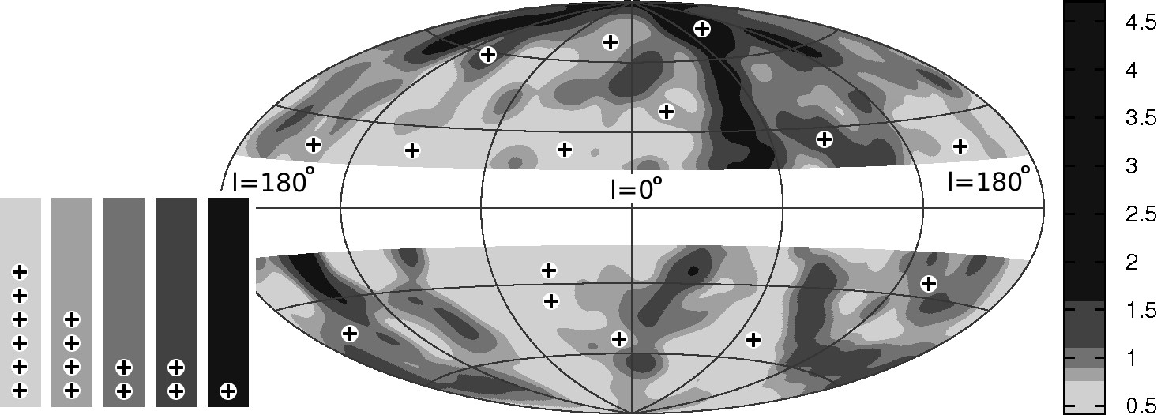}

\caption{\label{fig:skymap:example} Aitoff projection of the sky
(galactic coordinates $l$ and $b$; see also Fig.
\ref{fig:skymap:galaxies}) showing the relative integral CR flux above $40$ EeV
in the matter tracer model (grayscale) together with 15 fictitious CR events
(`+' symbols).
The gray bands are chosen such that each band contains 1/5 of the total
flux, with darker bands indicating a larger flux. Inset: distribution of the 15 CR events over the five bands.
}
}

The idea of the method is illustrated in Fig.~\ref{fig:skymap:example},
which shows the actual flux distribution above 40 EeV predicted in the \modelsp
together with 15 fictitious, uniformly distributed CR arrival directions.
(We discuss how this figure is obtained in Sect.~\ref{sec:CRdistmodel}).
The contours
of equal flux are drawn in such a way that they divide the sky into 5
areas, each containing an equal share of the total flux (1/5 in this
case). This implies that events produced according to the \modelsp
should be distributed roughly equally over the five bands.
On the contrary, if the sources are distributed in a
different way, one generally expects deviations from
equipartition. For instance, in the case of an isotropic CR flux
most events will fall in the region with the smallest flux
(the lightest band in Fig.~\ref{fig:skymap:example}) just because this
region covers the largest area.
Comparing the distribution of observed events 
over the five bands to a model prediction then
allows one to test the underlying model.
This is illustrated in the inset of
Fig.~\ref{fig:skymap:example} with a histogram that shows the number of events in
every band. For this specific example one would
reject the matter tracer model as the
distribution of events over the five bands
deviates significantly from equipartition. 

The method described above can be formulated in a binless way.
This is an important advantage because binning
introduces ambiguity which complicates the statistical analysis.
Our numerical method starts by
generating a (continuous) skymap of expected flux.
We then extract
flux values at the positions of data events, thus obtaining a set of flux
values which we refer to as the ``data set''. Then a large number of
simulated events is generated by Monte Carlo event generation
from the model that is to be tested
(e.g., an isotropic distribution or the \model).
The flux values at positions of simulated events
form the (Monte Carlo) ``reference set''.
Then a statistical test is performed
to determine whether the data and
reference sets are drawn from the same distribution.
This final step can be done with the
well-known Kolmogorov-Smirnov (KS) test. In this work we
also propose a modified version of the KS test, which we show to be
more sensitive than the standard KS test for the case at hand.
Both the KS test and its modification
do not require binning, rendering the whole procedure binless.
Furthermore, the method does not invoke scanning over unknown parameters
and hence avoids the use of penalty factors to estimate statistical
significances.

Although the main  purpose of this paper is to introduce our statistical test and analyze its potential
to infer properties of UHECR sources from future data, we have also applied it 
to the publicly available data from the Akeno Giant Air Shower Array (AGASA) and PAO
to test the \modelsp and an isotropic CR distribution.
As neither of the two hypotheses
can be ruled out, the present data give inconclusive results
which do not significantly add to the ongoing debate on the sources of UHECRs.
We find, however, that the proposed test offers a powerful
discrimination between source models with more CR events, and we believe
it is timely to present the method before such data become available.

The rest of this paper is organized as follows.
In Sect.~\ref{sec:modeling} we discuss our modeling of CR arrival
directions in the \model. This includes subsections on the modeling of
the source distribution, CR propagation, and on the
Monte Carlo event generation procedure.
Sect.~\ref{sec:tests} regards anisotropy tests. In this section we
discuss the test statistics considered in this work, 
and recapitulate some essential concepts in statistical hypothesis testing.
In Sect.~\ref{sec:powercomparison} we compare the statistical power
of the anisotropy test proposed in this work to other tests and we
consider its dependence on model parameters.
We apply our test
to data from AGASA and PAO in Sect.~\ref{sec:pval}.
We discuss our work in Sect.~\ref{sec:discussion}
and conclude in Sect.~\ref{sec:conclusion}.

\section{Modeling of CR arrival directions}
\label{sec:modeling}

In this work we will focus on two different models for
the distribution of CRs. The first is the \model, which assumes that
CR sources trace the distribution of matter and are numerous.
The second model (the ``isotropy'' model) predicts an isotropic distribution of CR events
over the sky, barring possible non-trivial detector exposure.
In this section we discuss our modeling of the CR arrival directions
in the \modelsp (the modeling of arrival directions in the isotropy
model being trivial).

\subsection{Source distribution; the KKKST catalog}
\label{sec:srcdist}
To estimate the distribution of CR arrival directions in the \modelsp
we must first model the distribution of their sources. The \modelsp presumes
that CR sources trace the large-scale distribution of matter in the Universe.
To our knowledge the most accurate way to model the large-scale
structure is by using galaxy catalogs obtained from extensive surveys.
There is however some ambiguity in this
procedure
because different catalogs lead to different reconstructed matter density
distributions. The reasons for this are twofold.
First, different types of galaxies exhibit somewhat different clustering properties
so that selection effects will affect the reconstructed matter distribution.
Second, each of the available catalogs has certain
drawbacks that limit the reconstruction accuracy. These include
incompleteness of the catalog, limited sky coverage, and uncertainties in
redshift determination.

For the present study we have considered two different galaxy catalogs:
the PSCz catalog \cite{Saunders:2000af} and a catalog that was recently compiled by 
Kalashev et al. \cite{Kalashev:2007ph}, hereafter referred to 
as the KKKST catalog.
The PSCz catalog contains $\EE{1.5}{4}$ galaxies up to 4 Gpc.
It has the advantage of accurate redshift determinations, but it is 
is incomplete and it does not cover all regions of the sky.
Catalog incompleteness may be taken into account by a selection function;
however this necessarily invokes some extrapolation. The incomplete sky coverage
is also a disadvantage because it necessitates either interpolating the matter density
over the left-out regions or a specialized treatment to include only
a specific part of the sky.

The KKKST catalog, compiled 
from the 2MASS XSC catalog \cite{Jarrett:2000me} and the (HYPER)LEDA database
\cite{2003A&A...412...45P}, contains
$\EE{2.1}{5}$ galaxies within 270 Mpc.
It is a complete sample up to 270 Mpc for galactic latitude $|b| > 15\degree$. 
The 2MASS XSC catalog is at present the most complete all-sky galaxy catalog.
Radial velocity measurements are however unavailable for the bulk of the galaxies,
so that one has to
rely on photometric redshift estimates (assuming galaxies are standard
candles in the infrared).\footnote{The
XSCz catalog (see \texttt{http://web.ipac.caltech.edu/staff/jarrett/XSCz/}) contains radial velocity
measurements for 27\% of the 2MASS galaxies. It
is however not (yet) publicly available.}
As discussed in Ref.~\cite{Jarrett:2004nk}, photometric redshift estimates 
lead to large uncertainties
in individual distance estimates but yield correct estimates when 
averaged over a sufficiently large number of galaxies.
For galaxy groups and clusters, the accuracy is estimated to be within $\sim$$20$\%
\cite{Jarrett:2004nk}.
Although the 2MASS XSC galaxy catalog is expected to lead to accurate reconstructed
matter densities at large distances,
where there are many galaxies,
it may not be reliable at smaller distances. For this reason
the KKKST catalog uses the (HYPER)LEDA database, which contains measured radial
velocities, at distances below 30 Mpc.

In this work we will mostly use the KKKST catalog to reconstruct the local matter density in the
Universe, choosing inaccuracy in
redshift determination over inaccuracies associated with catalog incompleteness.
(In Sect.~\ref{sec:powercomparison} we will also recompute some of our results
using the PSCz catalog to study the effect of the assumed galaxy catalog on the
power of the statistical tests proposed in this work.)
We find the inaccuracies due to photometric redshift estimates acceptable,
given the other uncertainties we are faced with. In particular, the uncertainties
in CR energies due to experimental resolution are of the order 20\%, which translates
to an uncertainty in viewing distance (and hence on the distance scale) 
$\gtrsim 30$\%.
To assess the differences between the PSCz and the KKKST catalogs,
we have compared the CR fluxes predicted by the \modelsp with the matter density
in the Universe reconstructed from both catalogs.
We find, reassuringly, that the flux distributions over the sky are
similar in the sense that the same physical structures may be recognized (see also below),
although some are brighter in one catalog compared to the other. 
The most important difference between the two is that the
KKKST catalog gives rise to stronger contrasts between under- and overdense regions.
This  also implies a stronger contrast in expected CR fluxes and hence
in larger differences between predictions from the \modelsp and an isotropic CR distribution.
Whether or not the true CR sources exhibit these relatively large contrasts is
at present unknown. Our approach is to adopt the KKKST catalog, and hence the 
relatively large density contrasts, at face value
and let current and future data decide whether or not the predictions derived
from it are accurate.

The KKKST catalog contains five objects that are closer than 5 Mpc. We remove
these from the catalog because
they do not represent a proper
statistical sample of the local structure, while they would outshine
all other sources due to their proximity.\footnote{This notably
excludes Centaurus A, with a distance of $\sim$$3.5$ Mpc  the nearest active galaxy, from our analysis.
The PAO data raise the
possibility that this galaxy is a source of
UHECRs \cite{Cronin:2007zz, Fargion:2008sp,  Gureev:2008bj,  Hardcastle:2008jw, Gorbunov:2007ja, Wibig:2007pf, Moskalenko:2008iz}. Other methods (e.g., small-scale anisotropy tests) are more appropriate to test this possibility.}

The distribution of galaxies as a function of distance
is shown in figure \ref{fig:srcdistD}. As can be seen in the figure, the distribution
is consistent with a uniform density 
at distance larger than a few ten Mpc, which reflects the completeness
of the sample. At smaller distances
there is an overdensity which we attribute to galaxies in the 
local supercluster. 
The global density of galaxies in the catalog is $n_0 = \EE{3.5}{-3}$ Mpc$^{-3}$.

\FIGURE{
\includegraphics[width=6.5cm, angle=270]{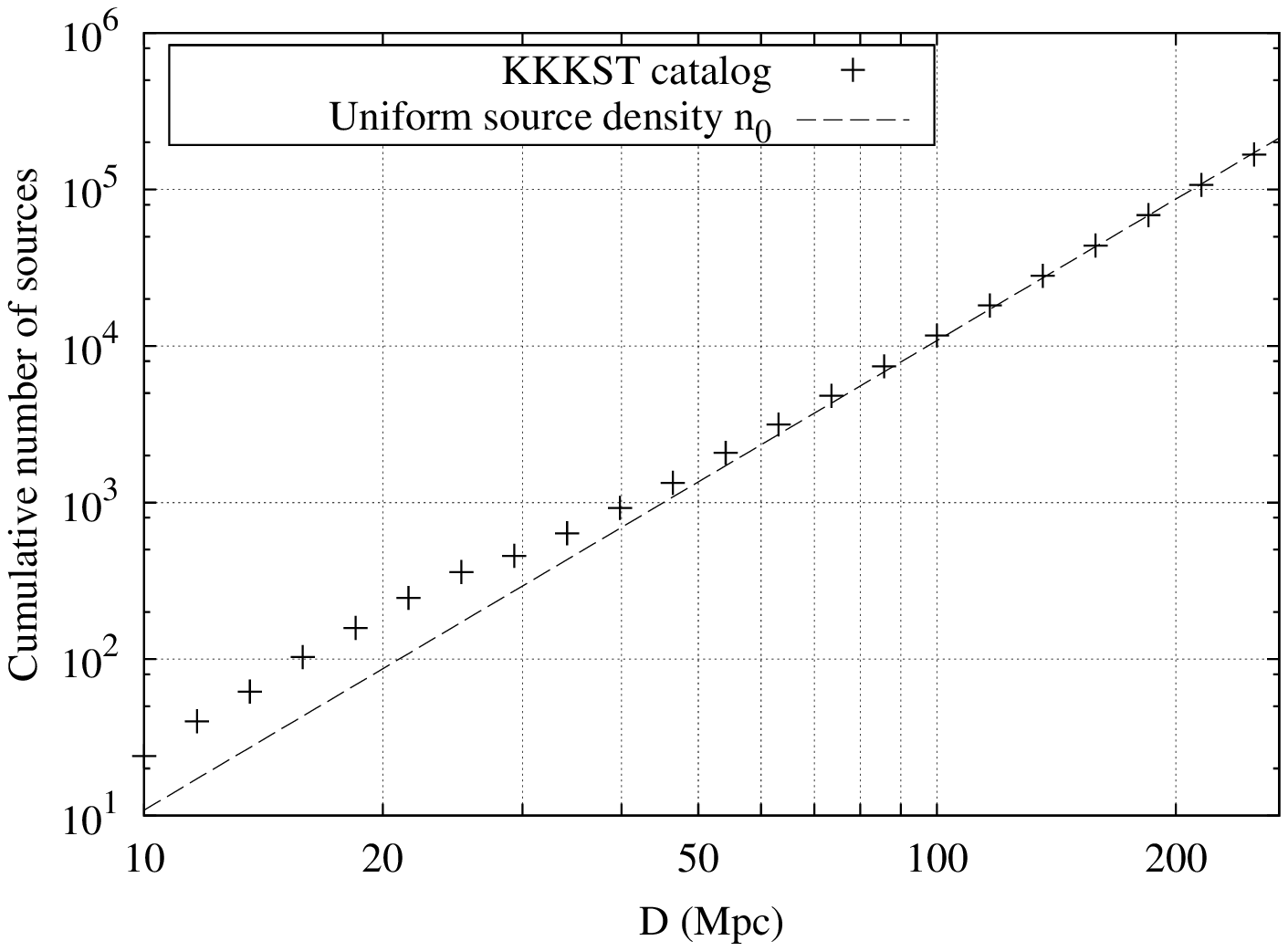}
\includegraphics[width=6.5cm, angle=270]{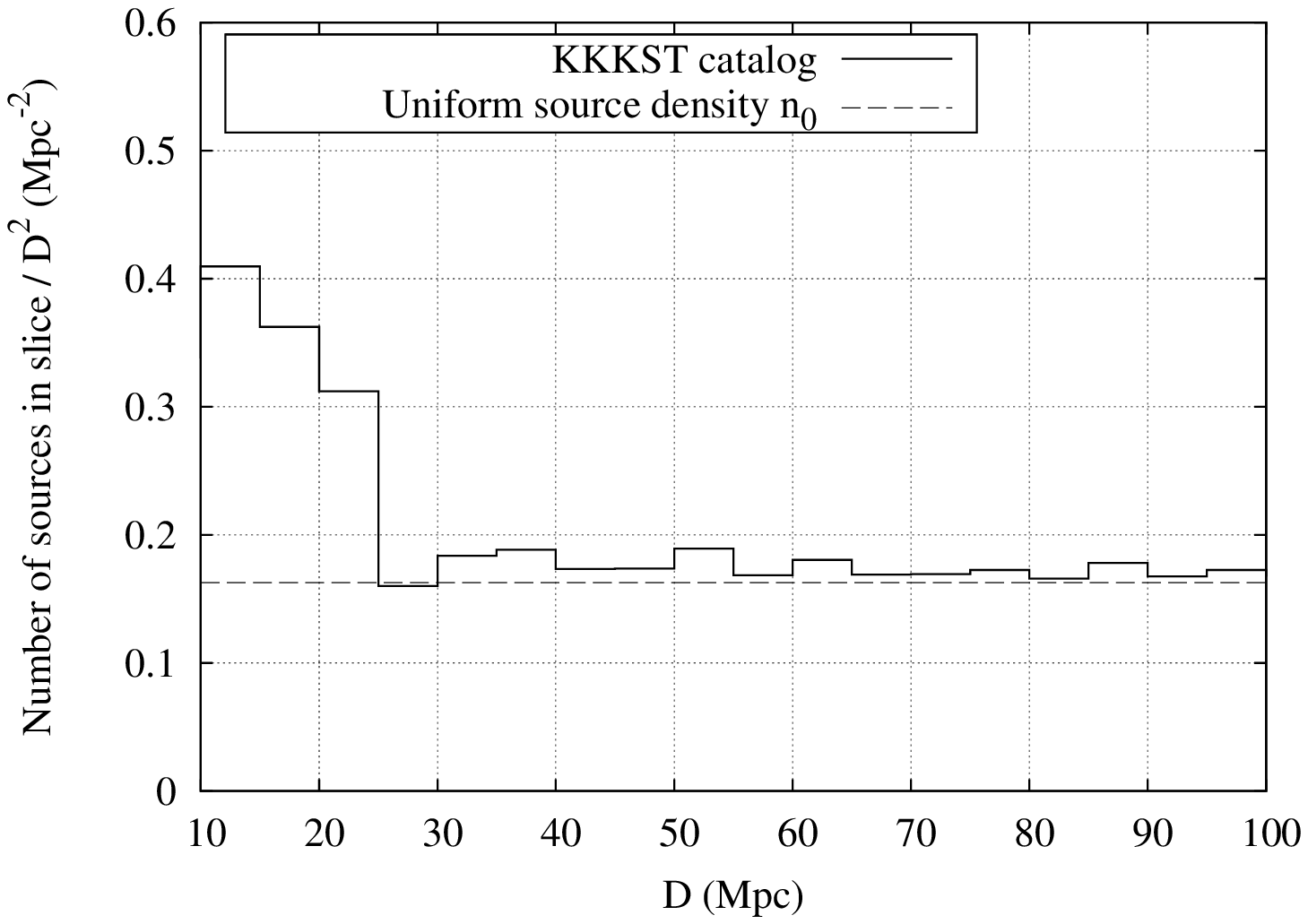}
\caption{\label{fig:srcdistD} Distribution of galaxies in the
KKKST catalog. Top panel: cumulative distribution as a function
of distance. Bottom panel: number of sources within a slice of
5 Mpc divided by the distance-squared. In both panels the data in the
catalog is compared to the expected distribution for a uniform
source distribution with density $n_0 = \EE{3.5}{-3}$ Mpc$^{-3}$.}
}

The absence of data at large distances and small galactic latitude (close to the galactic plane)
necessitates extrapolation. At distances beyond 270 Mpc, 
we will assume that sources are distributed uniformly.
We exclude the region within $15\degree$ of
the galactic plane from our analysis. 
To compensate for artificial boundary effects due to this
cut, we extrapolate the catalog data
in a straightforward manner
when computing average source densities just outside
the excluded region: for every galaxy with coordinates $l$ and $b$,
where $|b|<30\degree$, we add a mirror galaxy with coordinates
$l'=l$ and $b' = \pm 30\degree - b$ in the galactic strip. Here the
plus (minus) sign corresponds to $b>0$ ($b<0$).

In Fig.~\ref{fig:skymap:galaxies} we show the expected integral CR flux above
40 EeV in the \model, using the KKKST catalog to reconstruct
the local matter density in the Universe, together with the positions of some prominent
groups and (super)clusters of galaxies. (We discuss how this
figure is obtained in Sect.~\ref{sec:CRdistmodel}). In this figure, as in all figures 
in this paper showing model CR fluxes on the sky, the gray bands
are chosen such that they  contain 1/5 of the model flux each, with darker bands indicating
larger flux. (We would like to stress that this
division in a discrete number of bands is for presentation only.) Both far-away superclusters
and close-by galaxy groups can be recognized in the figure, the most prominent overdense region 
extending between the local (Virgo) and Centaurus superclusters.

\FIGURE{
\includegraphics[height=5.3cm]{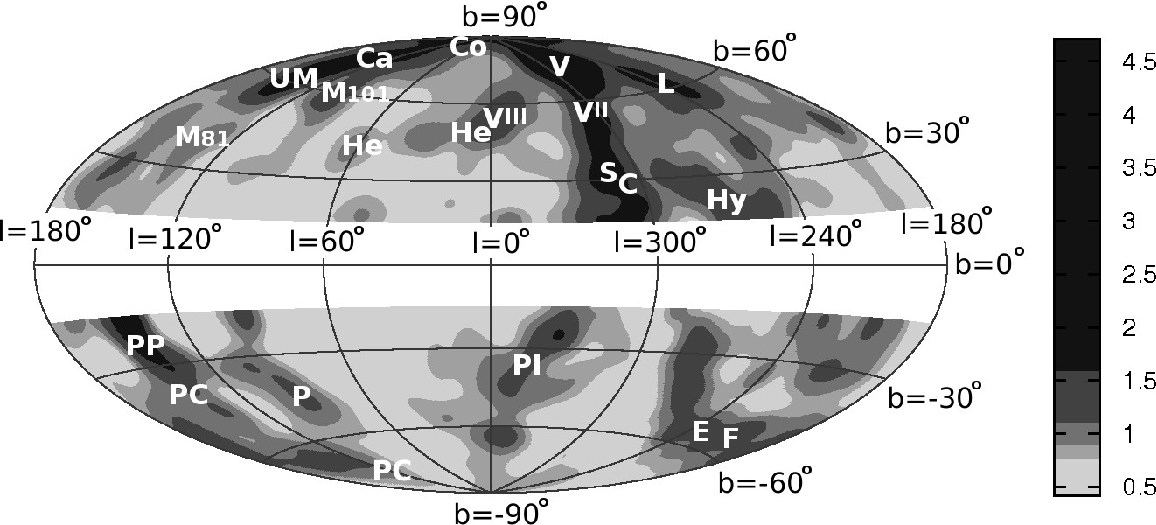}
\caption{\label{fig:skymap:galaxies} Aitoff projection of the sky
(galactic coordinates $l$ and $b$) showing the relative integral CR flux above $40$ EeV
in the matter tracer model (grayscale;  matter densities derived from the KKKST catalog)
together with the approximate
positions of some prominent
galaxy groups and (super)clusters.
C: Centaurus supercluster (60 Mpc);
Ca: Canes I group (4 Mpc) and Canes II group (9 Mpc);
Co: Coma cluster (90 Mpc);
E: Eridanus cluster (30 Mpc);
F: Fornax cluster (20 Mpc);
He: Hercules superclusters (140 Mpc);
Hy: Hydra supercluster (50 Mpc);
L: Leo supercluster (130 Mpc), Leo I group (10 Mpc), and Leo II group (20 Mpc);
M81: M81 group (4 Mpc);
M101: M101 group (8 Mpc);
P: Pegasus  cluster (60 Mpc);
PI: Pavo-Indus supercluster (70 Mpc);
PC: Pisces-Cetus supercluster (250 Mpc);
PP: Perseus-Pisces supercluster (70 Mpc);
S: Shapley supercluster (200 Mpc);
UM: Ursa Major supercluster (240 Mpc), Ursa Major North group (20 Mpc), and 
Ursa Major South group (20 Mpc);
V: Virgo cluster (20 Mpc);
VII: Virgo II group (20 Mpc);
VIII: Virgo III group (20 Mpc). The Pisces-Cetus supercluster
extends between the two indicated points.
Distances between parentheses are rough estimates.
}
}

\subsection{Source density}
\label{sec:srcdens}
The CR source density $n_{\rm src}$ is unknown, but bounded from below 
by the scarcity of observed doublets and triplets, 
and also by the fact that the sources of the observed
UHECRs cannot be very far away because of CR energy loss in the Universe.
These considerations lead to a (rough) lower bound 
$n_{\rm src} \gtrsim 10^{-5}$ Mpc$^{-3}$ \cite{Waxman:1996hp}.

The  CR source density  affects the distribution of CR
arrival directions through event clustering: if there are fewer
sources, one expects more doublets, triplets, etc.
In the limit of very many sources  the 
CR distribution becomes independent of source density,
as the probability to observe more than one CR from
any source is small regardless of the exact value of the source
density. We will refer to this situation as
the many-source regime.

The \modelsp considered in this work is built on
the assumption  of a large number of observable sources.
Since we model the distribution of CR sources
from the KKKST catalog, 
we should verify that the number density of our model sources
is sufficiently large
for the many-source regime to apply 
(if it were too small we would risk introducing spurious event clustering).
Assuming equal source luminosities, the number
of CR sources $N_{\rm src}$ may be estimated from the number of singlets $n_1$
and doublets $n_2$ \cite{Dubovsky:2000gv}:
\beqs
\label{eq:Nsrc}
N_{\rm src} \simeq \frac{n_1^3}{2 n_2^2} \, .
\eeqs
Demanding, somewhat arbitrarily, that
$n_2/n_1 = 0.1$ we find that $N_{\rm src} / n_1  \simeq 60$.
This translates to a typical source density $ \hat{n}_{\rm src}$, above which
the many-source regime applies:
\beqs
\hat{n}_{\rm src} = 10^{-4}  \lb \frac{N_{\rm ev}}{10} \rb\lb \frac{D}{100 \, {\rm Mpc}}\rb^{-3} \, {\rm Mpc}^{-3} \, ,
\eeqs
where $N_{\rm ev} \simeq n_1$ is
the total number of CR events
and $D$ denotes the maximum viewing distance (which is limited due
to interactions between CRs and the CMB).
For realistic 
numbers of events the source density in the KKKST catalog
$n_0 \gg \hat{n}_{\rm src}$ and the many-source regime
applies. Hence we can model the distribution of CR arrival directions
in the \modelsp from the KKKST catalog.

In Sect.~\ref{sec:powercomparison}
we will be interested in the relation between source density
and power of the statistical test proposed in this work.
As a matter of convenience we parameterize the source density 
by the density fraction
\beqs
f_{\rm src}  := \frac{n_{\rm src}}{n_0} \, .
\eeqs
The lower bound on the CR source density then implies that
$f_{\rm src} \gtrsim \EE{3}{-3}$.

\subsection{Cosmic-ray propagation}
\label{sec:CRprop}
In this section we consider modifications to the CR spectrum
due to redshift and interactions with the cosmic microwave
background (CMB). 
Throughout this work we will assume that CRs are protons,
the composition of CRs at the highest energies being unclear \cite{Unger:2007mc, Glushkov:2007gd, Abbasi:2004nz}.
CR energy loss in the Universe may be described by the 
following differential equation
(see, e.g., Ref.~\cite{Berezinsky:2002nc})
\beqs
\label{eq:Eg:diffeq}
\frac{1}{E} \frac{dE}{dz} = \frac{1}{1+z} + \frac{(1+z) \beta_0 ((1+z)E)}{H (z)}  \, ,
\eeqs
where $E$ is the proton energy, $z$ denotes the source redshift, $H(z)$ is defined below, and
$\beta_0$ denotes the inverse energy loss time (at present epoch)
due to CR-CMB interactions. 
We include CR energy loss due to photopion production and
electron-positron pair production. For these processes we use an
approximation for $\beta_0$, valid for energies above $10^{17.5}$ eV, 
which may be found in the appendix of Ref. \cite{Koers:2008hv}.
This simplified treatment 
does not account for the stochastic nature of CMB-CR interactions
nor for the limited energy resolution in the detector (both effects lead to underestimation of the
mean free path; see, e.g., Ref.~\cite{Kachelriess:2007bp}), 
and it neglects interactions with the extragalactic background light
(which overestimates the mean free path; see, e.g., Ref.~\cite{Dermer:2008cy}).
In the light of
other limitations, notably the scarcity of  data at the highest energies, we consider 
these simplifications justified in modeling the arrival directions of CRs. 
We note that these simplifications do not affect the statistical test in itself.

Within the
$\Lambda$CDM concordance model, the redshift $z$ and proper 
distance $D$ are related as follows:
\beqs
D (z)
= \int_0^z  dz' \frac{c}{H(z')} \, ; \quad 
H(z) := H_0 \sqrt{\Omega_m (1+z)^3 + \Omega_{\Lambda}} \, ,
\eeqs
where  $c$ is the speed of light, $H_0$ denotes the present Hubble rate, and $\Omega_m$ ($\Omega_{\Lambda}$)
is the present density of matter (vacuum energy) in units of the critical density.
We use the concordance model values $H_0 = 72$ km s$^{-1}$ Mpc$^{-1}$, $\Omega_m = 0.27$, and 
$\Omega_{\Lambda} = 0.73$. 

Evolving eq. \eqref{eq:Eg:diffeq} forward in time (backward in redshift) yields the
energy of a proton on its trajectory from  source to Earth.
In figure \ref{fig:EE0} we show the
CR energy as a function of traversed distance for different injection (source)
energies $E_0$. The figure clearly shows the strong energy loss
of ultra-high-energy protons that gives rise to the GZK cutoff in the
CR energy spectrum.

\FIGURE{
\includegraphics[width=6.5cm, angle=270]{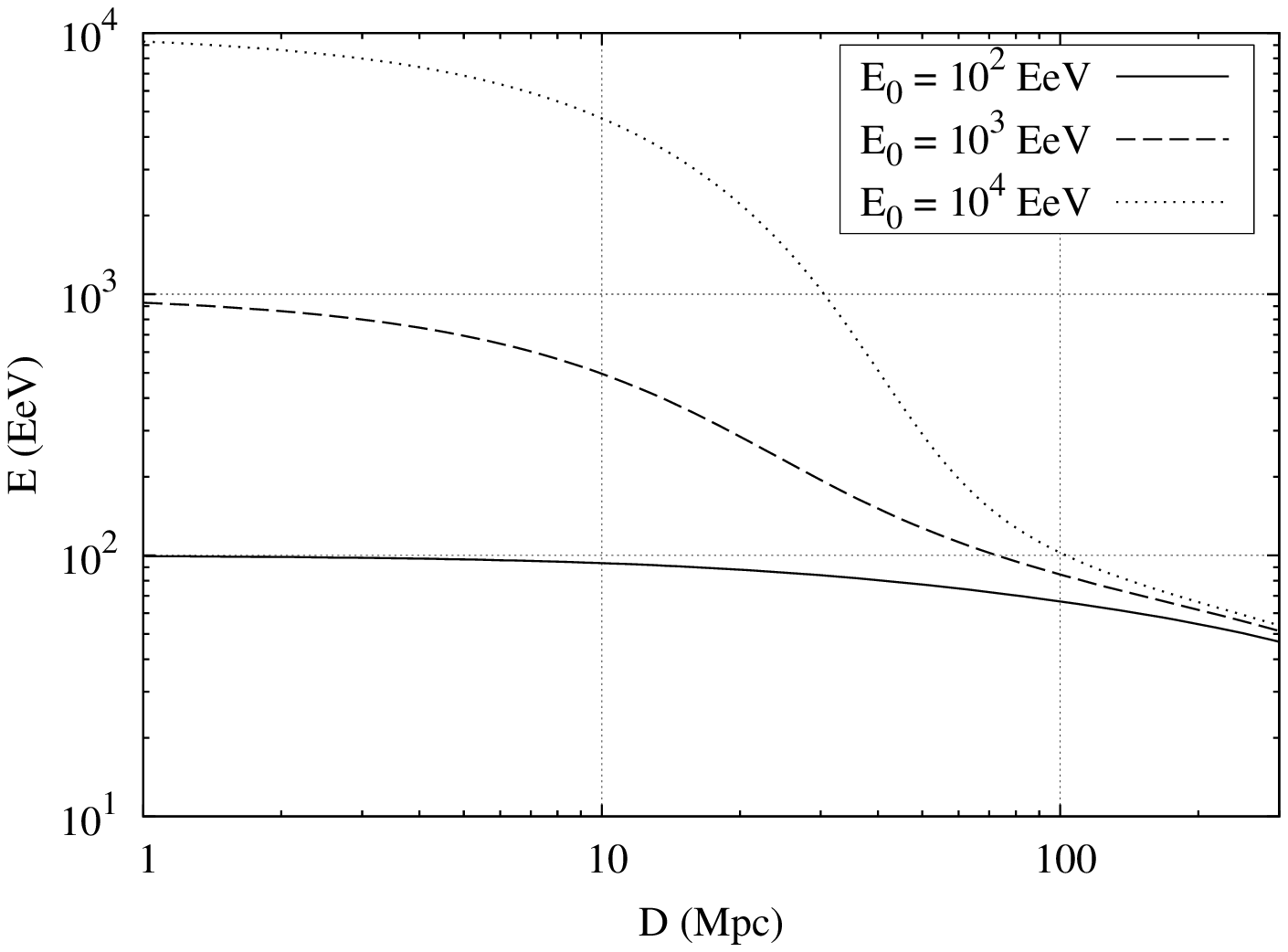}
\caption{\label{fig:EE0} CR proton energy $E$ as a function of traversed distance $D$ for
three different injection energies $E_0$.}
}

The physical quantities relevant to the present analysis are easily expressed in terms
of a function $E_0 = E_0 (E, D)$ that gives the CR energy at the source
$E_0$ as a function of observed energy $E$ and source distance $D$.
We determine this function by numerically evolving eq. \eqref{eq:Eg:diffeq} backward
in time (forward in redshift). In terms of $E_0$
the integral flux $\Phi(E)$ from a single source at distance $D$ is:
\beqs
\label{eq:ptflux}
\Phi(E) = \frac{J^0 (E_0)}{4 \pi D^2 (1+z)}  \, ,
\eeqs
where $J^0 (E_0)$ is the integral luminosity above $E_0$ of the source. 
Summing up the contributions
of all sources, the diffuse CR flux  may be expressed as follows:
\beqs
\label{eq:diffflux}
\Phi^{\rm diff} (E) =   \frac{c n_{\rm src}}{4 \pi} \int_0^{z_{\rm max}} dz \, 
\frac{\ep(z) J^0 (E_0)}{H(z) (1+z)} \, ,
\eeqs
where $n_{\rm src}$ is the present source density, $z_{\rm max}$ is the maximum redshift
(we use $z_{\rm max}=5$ unless otherwise stated), and $\ep(z)$ parameterizes
source evolution (no evolution corresponds to $\ep(z) \equiv 1$). In this equation
we assume, as we do throughout this work, identical sources.

\FIGURE{
\includegraphics[width=6.5cm, angle=270]{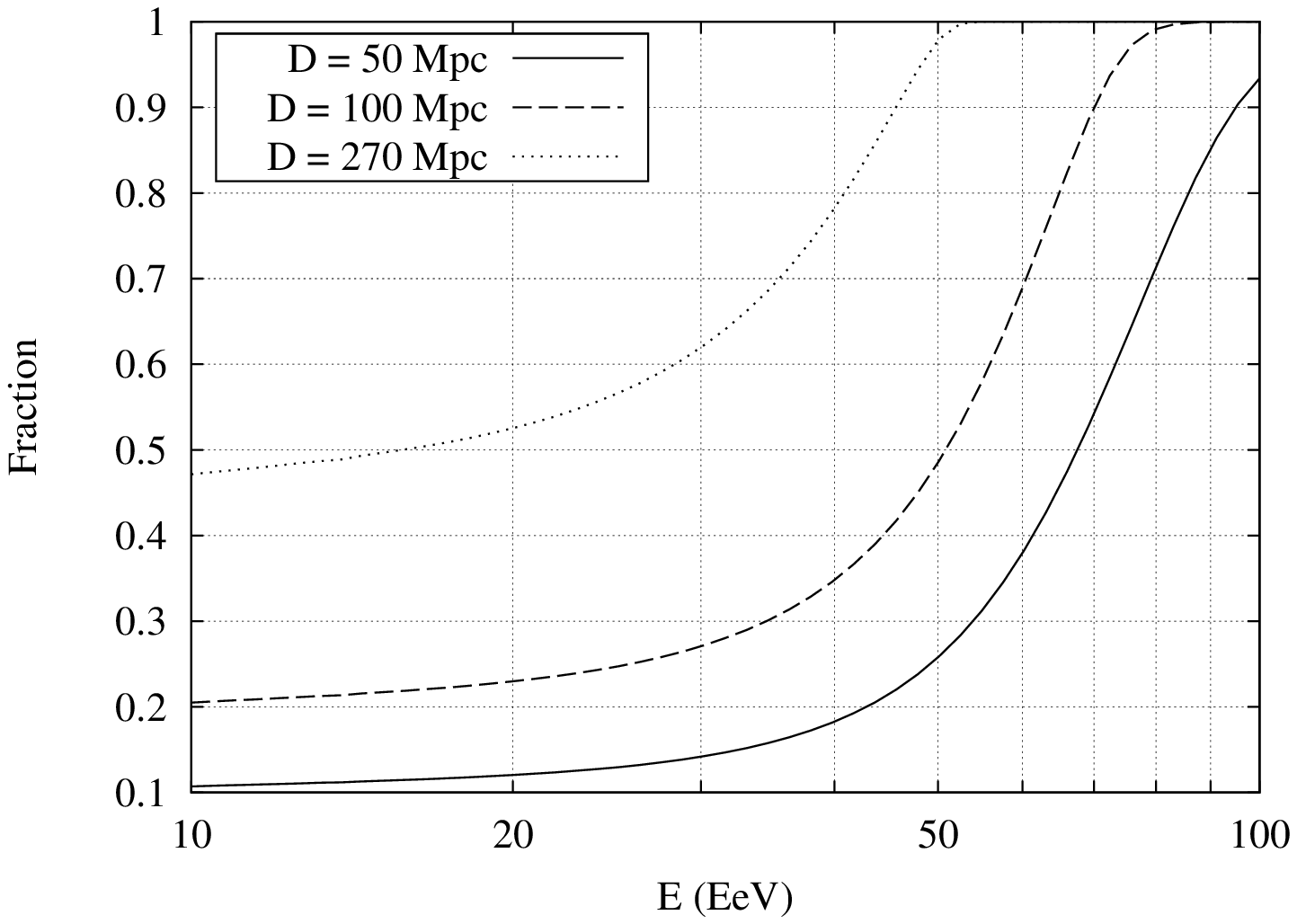}
\includegraphics[width=6.5cm, angle=270]{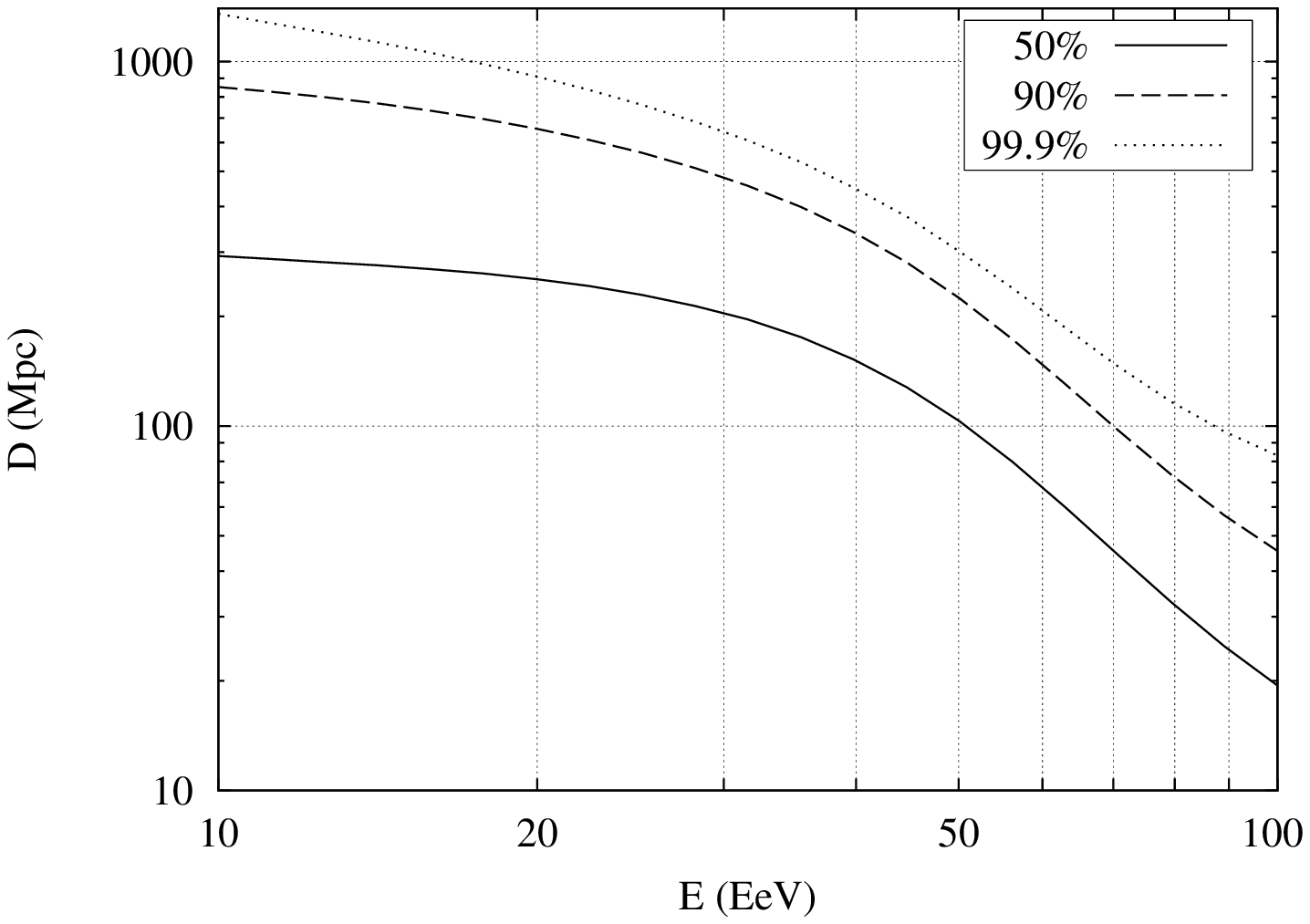}
\caption{\label{fig:gfrac} Top panel: fraction of diffuse (integral)
CR flux produced in sources
within distance $D$, as a function of energy.
Bottom panel: distance within which a given fraction of the CR flux
is produced as a function of energy.
In producing these figures we have
assumed an injection spectrum
$d N/dE \propto E^{-p}$ with $p=2.7$ extending to energies
much larger than 100 EeV, and no source evolution.}
}

\subsection{Cosmic-ray flux distribution in the \model}
\label{sec:CRdistmodel}
In this section we express
the CR flux from a direction on the sky 
in the \modelsp using the KKKST catalog and
the CR propagation model discussed in the previous subsections.
We are interested in the integral flux above some threshold
energy $E$. As the catalog is limited to a catalog distance $D$,
we first separate the total flux into two parts:
\beqs
\Phi (l,b,E)= \Phi^{<D} (l, b, E) + \Phi^{>D} (l, b, E) \, ,
\eeqs
where $\Phi^{<D}$ denotes the contribution
that arises from sources within the catalog distance $D$,
$\Phi^{>D}$ denotes the contribution from
sources at larger distance (which is
isotropic by assumption), and $l$ and $b$ are
galactic coordinates.
We parameterize the relative strength of the two components
as follows:
\beqs
\label{eq:defg}
g(E) = \frac{\int d \Omega \, \Phi^{<D} (l, b, E)}{\int d \Omega \, \Phi (l,b,E)} \, ,
\eeqs
which defines $g$ as the fraction of the total CR flux 
above energy $E$
produced by sources within distance $D$.
We compute this fraction from eq.
\eqref{eq:diffflux} by integrating up to redshift
$z_{\rm max}$ corresponding to $D$
and dividing the result by eq. \eqref{eq:diffflux} integrated to 
$z_{\rm max}=5$. This fraction is shown in the top panel
of Fig.~\ref{fig:gfrac}
for three values of $D$. In producing this figure we have assumed an injection spectrum
$d N/dE \propto E^{-p}$ with $p=2.7$  extending to energies
much larger than 100 EeV, and no source evolution.
As may be seen from the figure, the fraction of sources beyond 270 Mpc
contributing to the observed CR flux is essentially zero above $\sim$$50$ EeV.
In the bottom panel 
of Fig.~\ref{fig:gfrac} we show the effective CR viewing distance
as a function of CR energy. The three lines show the distances within which
50\%, 90\% and 99.9\% of the total flux are produced.

We model the contribution from sources within the KKKST catalog to
the diffuse CR flux as follows:
\beqs
\label{eq:def:PhiwithinD}
\Phi^{<D} (l, b, E) & = & \sum_i \Phi_i (E) w_i  \, ,
\eeqs
where $i$ enumerates the galaxies in the catalog,
$\Phi_i (E)$ is the integral flux above $E$ from galaxy $i$
(computed with eq. \eqref{eq:ptflux}), 
and
\beqs
w_i  = \frac{\exp \lb -\lb \theta/\theta_s \rb^2\rb}{\pi \theta_s^2}  \, ,
\eeqs
is a weighting factor that we insert to introduce angular smearing,
$\theta$ being the angle between galaxy $i$ and the line of sight
and $\theta_s$ being the angular smearing scale.
The angular smearing procedure 
is introduced to avoid unphysical fluctuations due to the use of a catalog of
point sources, and to account
for CR deflections in the intergalactic magnetic fields
and for limited angular resolution in the detector.
We conservatively
adopt $\theta_s = 6\degree$ in
this work (see Ref.~\cite{Kashti:2008bw} for a detailed discussion).

In figure \ref{fig:skymaps} we show
the relative strength of the CR flux on the sky 
in the \model.  Here we have used the KKKST catalog to reconstruct
the local matter density in the Universe.
The grayscale indicates the value of the quantity
\beqs
\frac{\Phi (l, b, E)}{\overline{\Phi} (l, b, E) } =
 g \lb \frac{\Phi^{<D} (l, b, E)}{\overline{\Phi}^{<D} (l, b, E)} \rb + 1-g \, ,
\eeqs
where overlined quantities are averaged over the sky.
In this figure we consider three different threshold energies:
40 EeV, 60 EeV, and 100 EeV.
The gray bands are chosen such that each band contains 1/5 of
the total flux, with a darker gray indicating a larger flux. 
A comparison between the different panels 
shows that the contrast in the
expected flux strengths increases strongly with increasing
CR energy. This is due to the fact that the nearby structure becomes
more prominent with decreasing viewing distance.

\FIGURE{

\vspace{0.5cm}

\includegraphics[height=4.5cm]{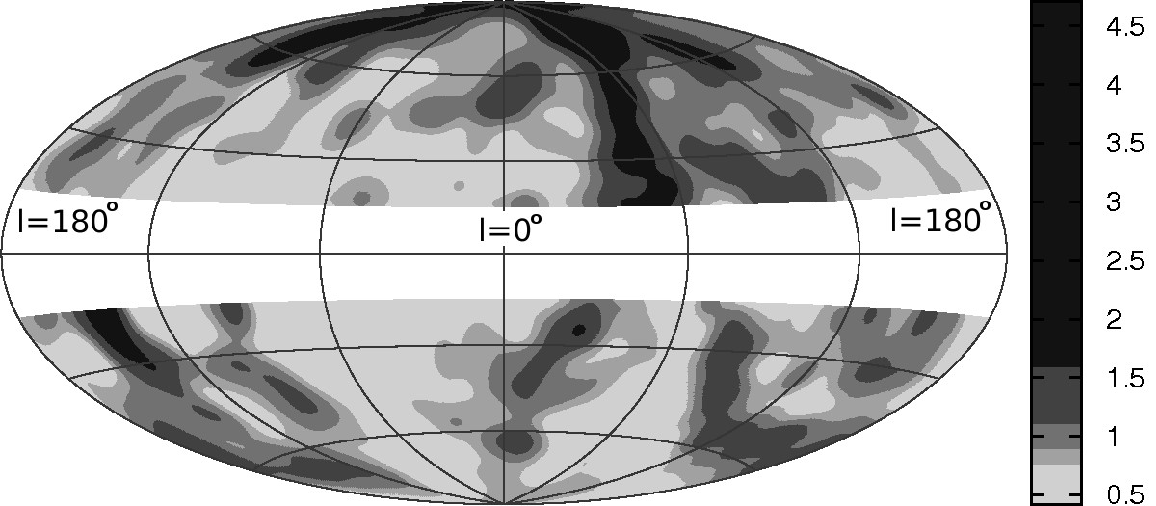}

\vspace{0.5cm}

\includegraphics[height=4.5cm]{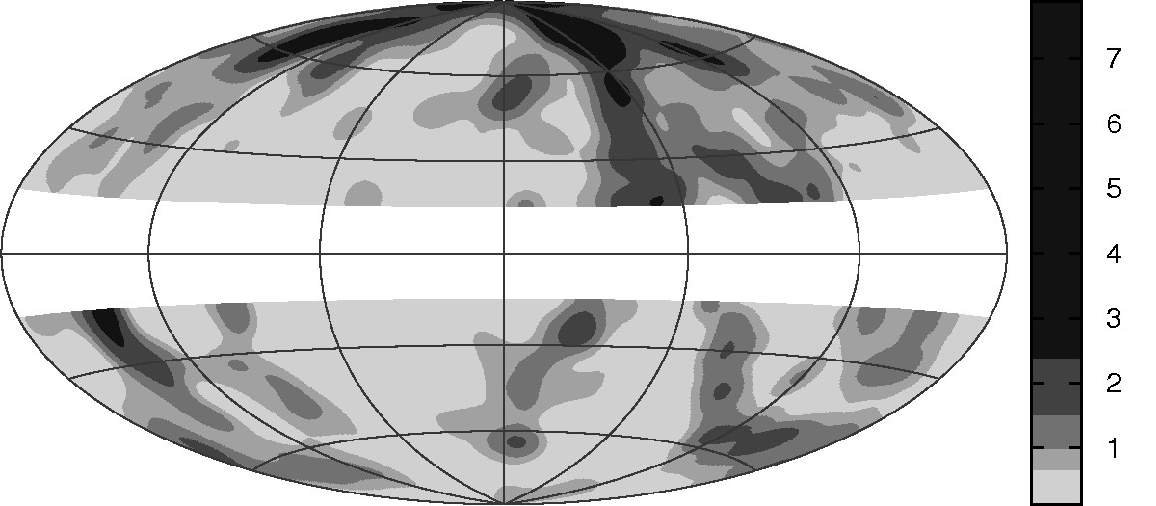}

\includegraphics[height=4.5cm]{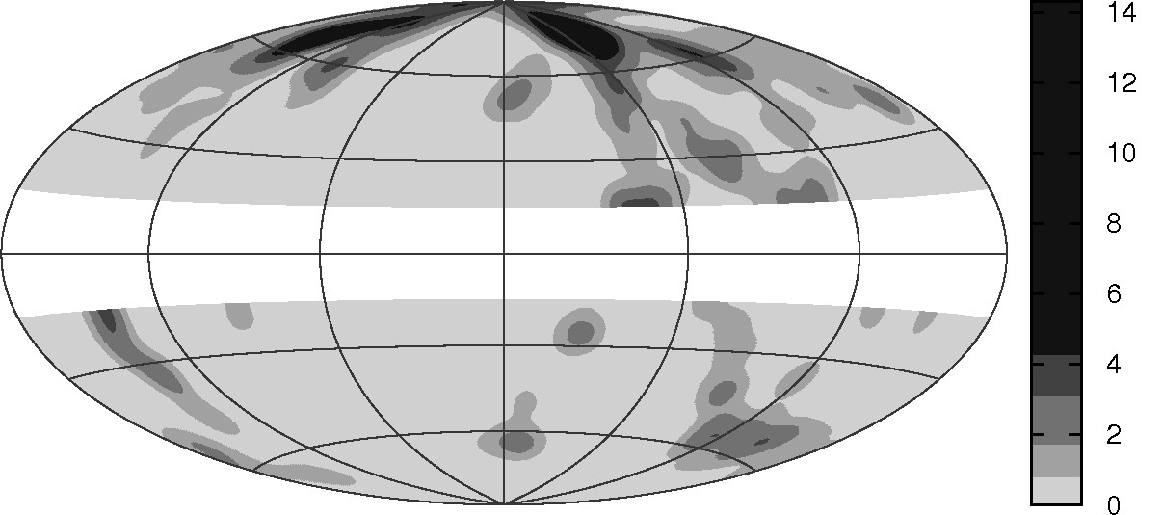}

\caption{\label{fig:skymaps} Aitoff projection of the sky
showing, from top to bottom, the relative CR flux
above $40$ EeV, $60$ EeV, and $100$ EeV
(same coordinates as in Fig. \ref{fig:skymap:galaxies}).
The grayscale indicates the relative integral flux in the \model,
averaged over $\theta_s=6\degree$, and with the local
matter density in the Universe derived from the KKKST catalog.
Each band contains one fifth of the total flux.}
}

\subsection{Event generation}
\label{sec:eventgen}
Here we discuss the Monte Carlo procedure employed in this work
to generate CR arrival directions (``events'').
The procedure relies on our modeling of the
galaxy distribution (discussed in Sect.~\ref{sec:srcdist}),
the density of sources (Sect.~\ref{sec:srcdens}),
CR energy loss (Sect.~\ref{sec:CRprop}), and on the detector exposure.

Event generation in the isotropy model is trivial: we determine arrival directions on the sky such
that their distribution  is proportional to the detector exposure.

In the \modelsp CR sources are selected from the KKKST galaxy catalog. 
Here the situation is complicated by the fact that
the catalog does not include sources beyond 270 Mpc.
In case the maximum viewing distance
is smaller than this
(which is the case for CR energies above $\sim$$50$ EeV), we proceed as follows.
First we select a galaxy from the KKKST catalog, where the probability $P_i$ to 
select galaxy $i$ is
\beqs
\label{eq:def:Pi}
P_i \propto  A_i \Phi_i(E)\, ,
\eeqs
$A_i$ being the relative  detector exposure in the direction of the event
and $\Phi_i(E)$ the expected flux from galaxy $i$ above energy $E$
(cf. eq. \eqref{eq:ptflux}). Once
a galaxy is selected, we generate an arrival direction around the
direction of the galaxy such that the  
distribution of the
angle $\theta$ between galaxy and arrival direction is
\beqs
\frac{d N}{d \theta} \propto \theta  \exp \lb - (\theta /\theta_s)^2 \rb \, ,
\eeqs
where $\theta_s$ is the smearing angle
(see Sect.~\ref{sec:CRdistmodel}).

If the maximum viewing distance is beyond 270 Mpc we 
add contributions derived under the \modelsp and the isotropy model 
(in doing so we assume that
CR sources are numerous and distributed uniformly  beyond 270 Mpc).
In this procedure we first
determine the fraction $g$ of the expected flux that comes
from within 270 Mpc
from eq. \eqref{eq:defg} (see also Fig.~\ref{fig:gfrac}, top panel).
We then randomly generate an event  following either the isotropy model (with probability $1-g$)
or the \modelsp (with probability $g$).

In Sect.~\ref{sec:powercomparison} we  will be interested in the effect of source density
on the statistical power of the test proposed in this work.
In the case of a CR source density smaller than the density of galaxies in the KKKST catalog
(i.e. $f_{\rm src} < 1$), we generate a reduced catalog with the desired source density 
by randomly selecting a fraction $f_{\rm src}$ of
galaxies from the full catalog. We then generate events from this 
reduced catalog 
following the procedure described above.

\section{Anisotropy tests}
\label{sec:tests}

\subsection{The contrast parameter}
\label{sec:chi}
The anisotropy test proposed in this work is based on the distribution
of a statistical parameter within a set of CR events.
We choose this parameter, which we denote as $\chi$, 
to be equal to the expected flux from sources within the catalog distance 
in the \model, modulated with detector exposure:
\beqs
\label{eq:def:chi}
\chi = A(l,b)\,  \Phi^{<D} (l, b, E) \, ,
\eeqs
where $A$ denotes the relative detector exposure, $\Phi^{<D}$ is defined in
eq. \eqref{eq:def:PhiwithinD}, $l$ and
$b$ are the (galactic) coordinates of the CR event, and
$E$ is the threshold energy.
Note that a choice of galaxy catalog is implicit in eq. \eqref{eq:def:chi}
via eq. \eqref{eq:def:PhiwithinD}; we
 use the KKKST catalog
unless stated otherwise.
Because $\chi$ is a measure of the
matter over- or underdensity in a given direction, we will refer to it
as the contrast parameter.
For the statistical tests considered in this work
the normalization of $\chi$ is arbitrary.

\FIGURE{
\includegraphics[width=6.5cm, angle=270]{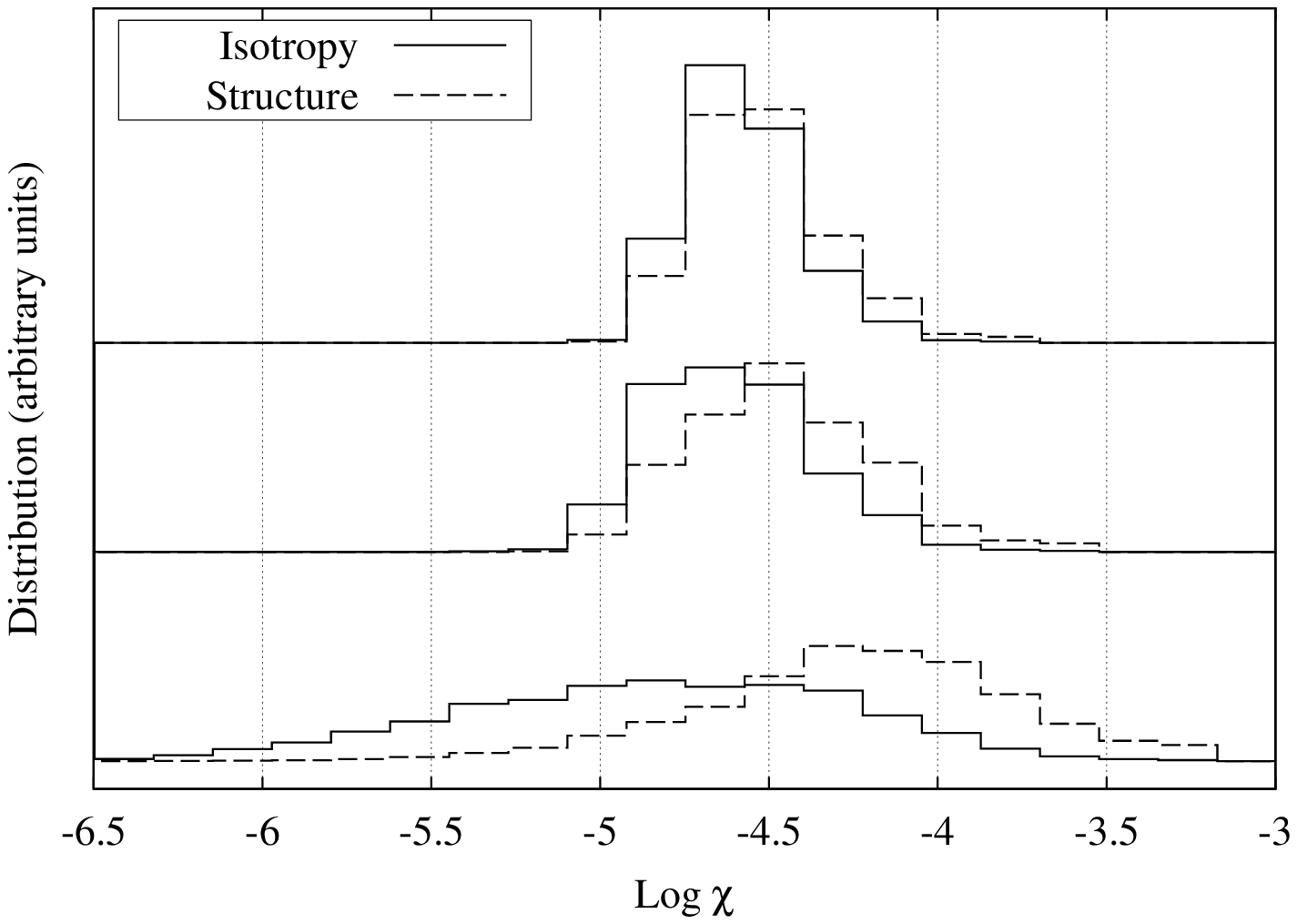}
\caption{\label{fig:chidist}Distribution of the contrast parameter
$\chi$  in the 
isotropic model (``Isotropy'') and the \modelsp (``Structure'')
for CR threshold
energies 40 EeV, 60 EeV, and 100 EeV (from top to bottom).
 Recall that  we use the KKKST catalog
to reconstruct the local matter density in the Universe
in the \model.
}
}

In figure \ref{fig:chidist} we show the distribution of the contrast parameter
$\chi$ when CR events are distributed isotropically and
in the \modelsp for three different CR energy thresholds. The figure
clearly shows that
$\chi$ takes preferentially larger values when events
follow structure, as expected.
This effect, which  is most pronounced at the highest energies,
underlies the statistical test proposed in the following section.

\subsection{Test statistics}
\label{sec:TS}
The statistical tests considered in this work are based on
test statistics, real-valued numbers that characterize a set of
CR events.
In this section we introduce the four test statistics
considered in this study.
We will discuss the methodology of hypothesis testing
in the following subsection.

First we consider the standard one-dimensional KS test statistic $D$ applied
to the contrast parameter $\chi$:
\beqs
\label{eq:defts:D}
D = \mathop{{\rm max}}_{-\infty <  \chi < \infty} \left[ \, \left| \mathcal{C}_{\rm test} (\chi ) - \mathcal{C}_{\rm ref} (\chi ) \right| \, \right]\, ,
\eeqs
where  $\mathcal{C}_{\rm test} (\chi )$ denotes the cumulative distribution of $\chi$
in the test data set and $\mathcal{C}_{\rm ref} (\chi )$ denotes a reference cumulative
distribution (all cumulative distributions will be assumed to be normalized). This reference distribution may be derived in the
model that is to be tested, but it may also be derived in an alternative model.

Second, for the purpose of this study we propose a novel
test statistic $Y$:
\beqs
\label{eq:defts:Y}
Y = \mathop{{\rm max}}_{-\infty < \chi < \infty} \left[ \mathcal{C}_{\rm test} (\chi) - \mathcal{C}_{\rm ref} (\chi) \right]
\, \,  - \mathop{{\rm max}}_{-\infty < \chi < \infty} \left[ \mathcal{C}_{\rm ref} (\chi) - \mathcal{C}_{\rm test} (\chi) \right] \, ,
\eeqs
which represents a small modification of the
standard KS test statistic $D$. This statistic is 
also similar to the Kuiper test
statistic, which is the sum of the two terms in eq. \eqref{eq:defts:Y}. 

In order to compare the sensitivity of the $D$- and $Y$-tests proposed here
we also consider two test statistics that were discussed in the literature
before.
The first of these is the two-dimensional KS test applied
to the galactic coordinates $l$ and $b$ of the CR events.
We use the implementation of this test
proposed in Ref. \cite{1987MNRAS.225..155F}, which is based
on earlier work in Ref. \cite{1983MNRAS.202..615P}. 
The procedure 
put forward in Ref. \cite{1987MNRAS.225..155F} yields a quantity $F$
that approximates the largest difference between the reference and test
distributions in the plane and over the four independent ways data can be
accumulated in two
dimensions:\footnote{Note that there are only two ways of accumulating data
in the one-dimensional
case, which are equivalent.}
\beqs
\label{eq:defts:F}
F \simeq \mathop{{\rm max}}_{0\degree < l < 360\degree, \, \, 15\degree < |b| < 90\degree, \, \, Q = 1\ldots 4} \left[ \, \left| \mathcal{C}^Q_{\rm test} (l, b) - \mathcal{C}^Q_{\rm ref} (l, b) \right| \, \right]\, ,
\eeqs
where $Q$ denotes one of the four quadrants, with respect to
a sampling point, in which data can be accumulated. 
Eq. \eqref{eq:defts:F} becomes an equality in the limit of an
infinite number of sampling points.
Due to computational constraints, we only consider a limited number of sampling
points, namely those points contained in the test data set.

Finally we will also consider the test statistic 
that was recently proposed by Kashti and Waxman
in Ref.~\cite{Kashti:2008bw}, who found it to be
more sensitive than methods based on the two-point correlation
function or the angular power spectrum.
We will denote this statistic as $X$ ($X_C$ in Ref. \cite{Kashti:2008bw}). The statistic is
obtained after binning the sky and comparing the number of events
per bin in the test data set to two independent model sets:
\beqs
\label{eq:defts:X}
X = \sum_i \frac{(N_{i, \, \rm test} - N_{i, {\, \rm ref}}) (N_{i, \, {\rm alt}} - N_{i, {\, \rm ref}})}{N_{i, {\, \rm ref}}} \, ,
\eeqs
where $N_{i, \rm test}$ denotes the observed number of events in bin $i$ in the test data;
$N_{i, {\, \rm ref}}$ and $N_{i, {\, \rm alt}}$ denote the average
numbers of events in bin $i$ in a reference model set and an alternative model set,
respectively; and $i$ enumerates the bins.

\subsection{Statistical power, significance, and $p$-values}
\label{sec:statistics}
In this section we recall the definition and interpretation of 
statistical power, significance, and $p$-values.

Our aim will be to test, i.e., to rule out or retain,
a particular null hypothesis $\Hnull$ (e.g., ``CR events are distributed
isotropically'').
Following the traditional frequentist approach put forward
by Neyman \& Pearson \cite{NP}, one may distinguish two types of errors
in this judgment:
\beqs
\nn \textrm{Type I error:} && \textrm{we reject  $\Hnull$ while it is true;} \\
\nn \textrm{Type II error:} && \textrm{we do not reject $\Hnull$ while an alternative hypothesis is true.}
\eeqs
The probabilities for the two types of error to occur are denoted as $\alpha$ and $\beta$, 
respectively. The first probability, $\alpha$, is known as the statistical
significance. In a frequentist sense it is just
the fraction of false rejections in many tests.
Note that $\alpha$ bears no reference to
alternative hypotheses. The statistical power, defined as $P=1-\beta$, 
is a measure of the ability of a given test
to discriminate between the null hypothesis and an alternative hypothesis;
it is computed for a particular alternative hypothesis.

The statistical significance and power are readily quantified for
hypothesis tests that are based on the value of a single test
statistic $t$. The tests considered in this work fall in this category.
In these tests one rules out the null hypothesis
when $t$ falls in some critical range, which has been specified a priori. 
The significance
$\alpha$ is then equal to the probability that $t$ falls inside the critical range
under $\Hnull$; the power of the test is the probability that $t$ falls
inside the critical range when the alternative hypothesis $\Halt$ is true.
The significance can be determined from the distribution of $t$ under $\Hnull$,
denoted as $\Dnull (t)$, which may be known analytically or sampled by
Monte Carlo event generation. Likewise, the power of the test is obtained from
$\Dalt (t)$, the distribution of $t$ under $\Halt$.
As an example consider the case that
large values of $t$ are incompatible with $\Hnull$. 
One would then decide to rule out
$\Hnull$ when $t_{\rm obs}$ is in the critical range
$(\hat{t},\infty)$ where $\hat{t}$ is a predetermined critical value.
The significance $\alpha$ of this test is then given by the integral
of $\Dnull (t)$ over the interval $(\hat{t}, \infty)$ and the power by the integral 
of $\Dalt (t)$ over $(\hat{t}, \infty)$. In practice one generally determines $\hat{t}$ from
a certain required significance $\alpha$.

Different from statistical power $P$ or
significance $\alpha$, the $p$-value is a data-dependent quantity. It
gives a figure of merit for the compatibility of a given measurement
and the null hypothesis. More precisely, it is
the probability that, under  $\Hnull$,  the outcome of an
experiment is \emph{at least as extreme} as the observed outcome.
Although small $p$-values indicate greater incompatibility with $\Hnull$, the
$p$-value has no precise frequentist interpretation.
In the case of a hypothesis test based on a single test statistic,
the $p$-value of a given measurement
can be determined by integrating $\Dnull (t)$ over the range $R$ of $t$-values that are 
equally or less compatible with the null hypothesis:
\nn
\beqs
\label{eq:defp}
p = \int_R \Dnull (t) \, dt \, .
\eeqs
In the exemplary case that large values of $t$ correspond to poor
agreement with the data, $R = (t_{\rm obs}, \infty)$, where
$t_{\rm obs}$ is the value of the test statistic corresponding to the observed data.

For the $X$- and $Y$-test statistics introduced in the previous section,
there is an important subtlety in choosing
the parameter range $R$ that covers values of the test statistic
``at least as extreme'' as a particular value $t_{\rm obs}$. This is because
for these test statistics
both very small and very
large values indicate disagreement between the data and $\Hnull$.
As there is no a priori reason to reject $\Hnull$ only for
very high or for very low values of the test statistic, 
we define $R$ as a combination of two disconnected parts that
cover both very small and very large values of $t$ (two-sided interval).
We choose these regions
symmetrically such that they contain equal fractions of
the test statistic distribution.
This procedure is equivalent to defining the $p$-value as follows:
\beqs
\label{eq:def:twosidedp}
p =  2 \int_{R'} \Dnull (t) dt \, ,
\eeqs
where by definition $R' = (t_{\rm obs}, \infty)$ when $t_{\rm obs}$ is higher than the average
value and  $R' = (-\infty, t_{\rm obs})$ when $t_{\rm obs}$ is smaller.

An insightful way to relate $p$-values to $\alpha$-probabilities is as follows.
Equations \eqref{eq:defp}
and \eqref{eq:def:twosidedp}
define a map $t_{\rm obs} \mapsto p_{\rm obs}$. Therefore we can 
represent a measurement by its $p$-value rather than its test statistic $t$. Under
$\Hnull$ the distribution of the $p$-value (over many experiments) is, by definition, uniform
between 0 and 1. We can thus rule out the null hypothesis, with significance
$\alpha$, when we determine experimentally that $p_{\rm obs}<\alpha$, where
$p_{\rm obs}$ is the $p$-value associated with a measurement via
eq. \eqref{eq:defp} and $\alpha$ is a \emph{predetermined} parameter.
As a note of warning we would like to stress that it is inconsistent
to interpret $p_{\rm obs}$ itself as a statistical
significance, which amounts to adjusting $\alpha$ to $p_{\rm obs}$ a
posteriori.


\section{Analysis of statistical power}
\label{sec:powercomparison}
In this section we consider the statistical power 
to rule out the ``isotropy'' model (CR events are distributed isotropically)
when the \modelsp holds.  This power is a measure of the
ability of the test to discriminate between the two models.
We reiterate that, within the \model, we reconstruct 
the local matter density in the Universe from the KKKST catalog unless
stated otherwise.
We will, in an obvious notation, refer to the
four hypothesis tests 
discussed in Sect.~\ref{sec:TS}
as the $D$-, $Y$-, $X$-, and $F$-tests.
The results in this section are based on
simulated event sets; we consider AGASA and
PAO data in the next section.

\subsection{Method}
\label{sec:powercomparison:method}
We compute the statistical power to reject an isotropic distribution
when the \modelsp is true
from the distributions of the test statistics
under both models (see section \ref{sec:statistics}).
We sample these distributions by Monte
Carlo event generation (see section \ref{sec:eventgen}). 
In this process we generate
$N_{\rm MC}$ independent test event sets that each
contain $N_{\rm ev}$ events.
For every simulated event we compute the contrast parameter $\chi$ from
eq. \eqref{eq:def:chi}.
We then compute
the values of the test statistics from eqs.  
\eqref{eq:defts:D}$-$\eqref{eq:defts:X}
for every simulated test set.
In these equations $\mathcal{C}_{\rm ref} (\chi)$, $\mathcal{C}_{\rm ref} (l,b)$ and
$N_{i, {\, \rm ref}}$ are computed from an isotropic reference event set,
while $N_{i, {\, \rm alt}}$
is obtained from an alternative model event set generated 
in the \model.
These reference and alternative model
event sets contain $N_{\rm ref}$ ($\gg N_{\rm ev}$) events.
Because of the stochastic nature of our routine (Monte Carlo
event generation),
there is some numerical uncertainty in our results.
We have verified
that the results presented in this section are accurate
within a few percent.

If the assumed CR source density is smaller than
the density of galaxies in the KKKST catalog, we use $N_{\rm SR}$ different
source realizations (reduced catalogs) to generate the $N_{\rm MC}$ test event
sets under the assumption that CR sources trace the distribution of matter.
Note that we compute the contrast parameter $\chi$ from the full
catalog also in this case because in practice we would not know
which fraction of the galaxies in the catalog are the actual CR sources.

\subsection{Results}
We are now in the position to compute the statistical power of the
hypothesis tests under consideration and investigate the dependence
on the different parameters.
To estimate the importance of event number and threshold energy
we will consider three scenarios:
\beqs
\nn \textrm{I}: && \textrm{100 CR events with energies above 40 EeV; }\\
\nn \textrm{II}: && \textrm{19 CR events with energies above 57 EeV} \\
\nn && \textrm{(corresponding to the PAO data with $|b|>15\degree$)}; \\
\nn \textrm{III}: && \textrm{10 CR events with energies above 100 EeV}.
\eeqs
Table \ref{table:power:DYXF} gives the statistical power
of the four hypothesis tests under consideration to rule out 
an isotropic distribution when the \modelsp is correct.
These results correspond to the following values for the
model parameters:
source density fraction $f_{\rm src}=1$, injection spectral index
$p=2.7$, maximum injection energy $E_{\rm max} \gg 10^{20}$ eV (the exact
value is unimportant in this regime), and smearing angle $\theta_s = 6\degree$.
Furthermore we have assumed no source evolution, and adopted 
the PAO detector exposure \cite{Sommers:2000us}.

\TABLE{
\begin{tabular}{| c | c | c c c c | c}
\hline
\,  Scenario  \,  &  \,  $\alpha$ \,& \, $P_D$ \, & \, $P_Y$ \, & \, $P_X$ \, & \, $P_F$ \, \\
\hline
\hline
I    &   0.01 & 0.31 & 0.33 & {\bf 0.58} &  0.068   \\
II   &   0.01 &  0.17 &  0.19 & {\bf 0.26} &  0.033  \\
III  &   0.01 & 0.55  & {\bf 0.61}  & 0.37 & 0.042 \\
\hline
I    &  0.05 &  0.55 & 0.58  & {\bf  0.79}  & 0.19  \\
II   &  0.05 &  0.37 & 0.41  & {\bf 0.48}  &  0.11 \\
III  &  0.05 &  0.78 & {\bf 0.84} & 0.64  & 0.13  \\
\hline
\end{tabular}
\caption{\label{table:power:DYXF}
Comparison of statistical power between
the four test statistics to rule out an isotropic
CR distribution when the \modelsp  (with local matter densities reconstructed from the
KKKST catalog) is correct.  We considered three different
scenarios for the number of events and minimum CR energy, PAO exposure, and two different
statistical significances: $\alpha =0.01$ and $\alpha =0.05$.
In producing these results we adopted a source density 
$n_{\rm src}=n_0=\EE{3.5}{-3}$ Mpc$^{-3}$ ($f_{\rm src} = 1$).
Furthermore we used $N_{\rm MC} = 10^5$ and $N_{\rm ref} = 10^5$ in our numerical
method.
Greatest powers are marked in boldface.}
}

As may be seen in this table, the
$Y$-statistic proposed in this work systematically yields a somewhat
higher power than the standard KS test statistic $D$. 
The power of the $F$-test is much smaller than the
power of the other tests in all three scenarios.
Comparing the powers of the $Y$- and $X$-tests, we find that the $X$-test
is superior in scenario I (large number of events at low energy) while
the $Y$-test performs better in scenario III (small number of events at high energies).
In the intermediate cases II (PAO data) the powers are comparable, with the $X$-test
performing slightly better.

We now focus on the $Y$-test, the most sensitive of the
two tests proposed in this paper, and
investigate how the statistical
power changes when we vary model parameters.
This is indicated in tables  \ref{table:power:f}
and \ref{table:power:varpars}.
The second column of these tables, labeled ``reference'', corresponds to the
reference values for the parameters which were also used in 
table \ref{table:power:DYXF}.
The other columns indicate how the statistical power changes when we 
vary one of these parameters, and when we use the PSCz catalog rather
than the KKKST catalog.

\TABLE{
\begin{tabular}{| c | c | c c c |  }
\hline
 Scenario &  Reference  & $ f_{\rm src} =10^{-1}$ & $ f_{\rm src}=10^{-2}$ & $f_{\rm src} = 10^{-3}$ \\
\hline
\hline
I  &  0.58 & 0.57 & 0.54 & 0.44 \\
II &  0.41 & 0.40  & 0.37 & 0.33 \\
III  & 0.84  & 0.82 & 0.69 & 0.49 \\
\hline
\end{tabular}

\caption{\label{table:power:f} Statistical power for the $Y$-test to rule out 
an isotropic distribution (with significance $\alpha=0.05$) when CR sources trace
the distribution of matter (as derived from the KKKST catalog) with varying source density.
In obtaining these results we used $N_{\rm MC} = 10^4$ ($N_{\rm SR}=10^2$),
and $N_{\rm ref} = 10^4$.  Note that these power estimates, as well as
those given in table \ref{table:power:varpars}, apply to
the PAO experiment.}
}

\TABLE{
\begin{tabular}{| c | c | c | c | c | c| }
\hline
 Scenario &  Reference   & $p=2.0$ & $E_{\rm max} = 200 \, {\rm EeV}$ & $\theta_s =10\degree $ &  PSCz\\
\hline
\hline
I   &  0.58 & 0.46 & 0.62 & 0.35 & 0.35 \\
II  &  0.41 & 0.33 & 0.45 & 0.25 & 0.15 \\
III &  0.84 & 0.74 & 0.95 & 0.62 & 0.59 \\
\hline
\end{tabular}

\caption{\label{table:power:varpars}  Statistical power for the $Y$-test to rule out 
an isotropic distribution (with significance $\alpha=0.05$) for different model parameters.
We used $N_{\rm MC} = 10^5$ and $N_{\rm ref} = 10^5$ in our numerical method.}
}

First let us consider the effect of the CR source density. As table
\ref{table:power:f} indicates,
the statistical power
of the test decreases with decreasing source density. This is due to the fact
that the distribution of $Y$ when sources trace the distribution of matter
becomes broader (being averaged over many source realizations), thus
increasing the overlap with the distribution of $Y$ when CRs are distributed
isotropically. This is illustrated in figure \ref{fig:Ydist} for scenario II.
As may be seen from the table, the statistical power convergences to an
asymptotic value as $f_{\rm src}$ approaches 1. This confirms our
earlier estimates that the many-source regime applies for a CR
source density equal to the density of galaxies in the KKKST catalog.

\FIGURE{
\includegraphics[width=6.5cm, angle=270]{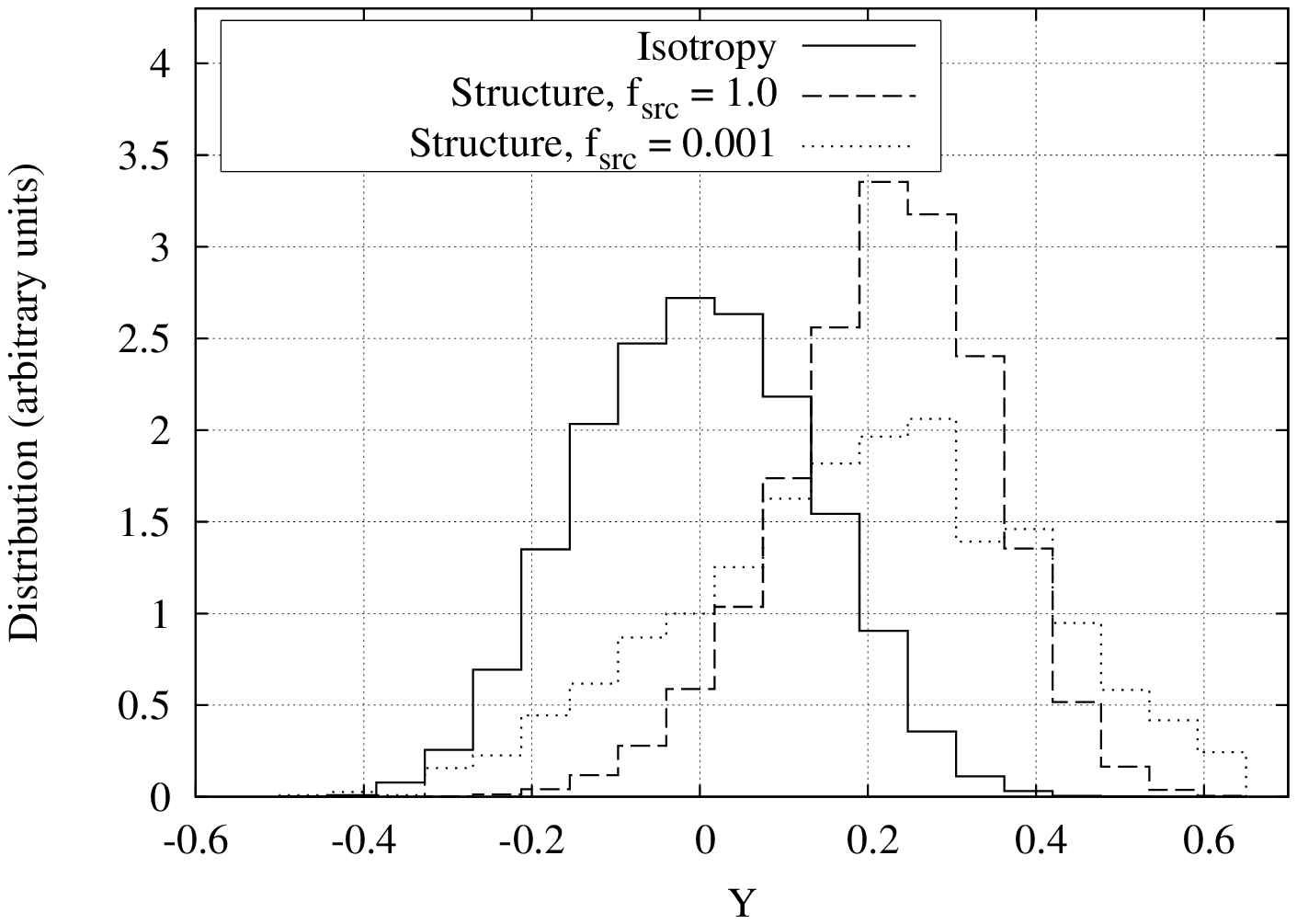}
\caption{\label{fig:Ydist} Distribution of $Y$ for test event sets
generated under the isotropy model (``Isotropy'') and 
under the assumption that CR sources trace the distribution
of matter (``Structure'') for scenario II (19 events above 57 EeV).
We considered $f_{\rm src}=1$ (many sources, 
hence ``Structure'' represents the \model),
and $f_{\rm src}=0.001$ (few sources).
We recall that we use the KKKST catalog
to reconstruct the local matter density in the
\model.
}
}

In table
\ref{table:power:varpars} we show how the statistical power varies
when we change different model parameters ($f_{\rm src}=1$ here).
First, we have varied the power-law index for the 
injected proton spectrum.  We observe that the statistical power decreases
when the injection spectrum becomes harder. This may be understood from the
fact that the effective viewing distance increases with a harder
injection spectrum. The increase in observable volume 
dilutes the contrast from nearby structure.
Second, we consider the effect of maximum injection energy.
Here we find that the statistical power increases when the
maximum CR energy upon injection is not much larger than the threshold energy.
This can also be understood in terms of viewing distance, with a lower $E_{\rm max}$ 
implying a smaller viewing distance and hence a stronger contrast.
Third, we have increased the smearing angle to $\theta_s=10\degree$. This 
obviously dilutes anisotropies and hence decreases the statistical power to
rule out an isotropic distribution when the \modelsp is true.
Fourth, we have investigated the statistical power of the proposed test when
the distribution of matter is derived from the PSCz catalog \cite{Saunders:2000af}
instead of the KKKST catalog.
Here we weight the flux of individual galaxies in the catalog with the
inverse of the selection function $\psi (r)$ (see Ref. \cite{Saunders:2000af}), adjusted
to $H_0 = 72$ km s$^{-1}$ Mpc$^{-1}$.
To avoid large fluctuations due to a few far-away sources, we cut the 
PSCz catalog at 270 Mpc and assume an isotropic distribution at larger
distance. This distance is chosen for comparison with the KKKST catalog.
For  the same reason we also cut the PSCz catalog
at $|b|<15\degree$.
We observe that the statistical power decreases substantially when
we replace the KKKST catalog by the PSCz catalog.
This is due to the fact that matter density contrasts are less
pronounced in the PSCz catalog, an issue which will be discussed
in Sect.~\ref{sec:discussion:X}.
Finally (not shown in the table), we find that source evolution
has little effect on the statistical power.
Even when we assume that CR sources trace strong AGN luminosity density evolution
as parameterized in Ref. \cite{1998MNRAS.293L..49B}, the results remain the same
within numerical uncertainties.
This is expected because high-energy CRs originate from sources in the local
Universe, rendering the effect of source evolution small.

We have compared the statistical powers of the $X$- and $Y$-test
in the first part of this section.
Qualitatively, we have found that the $Y$-test performs better at 
high energies and the $X$-test better at low energies.
The precise energy where the two tests are equally powerful depends
on many parameters, such as 
the experiment's field-of-view, the deflection angle,
and the assumed CR spectrum. Using reference values of the model parameters
we found that the $Y$-test is the most sensitive
in scenario III, while the $X$-test is the most sensitive
for scenarios I and II (see table  \ref{table:power:DYXF}).
We have verified explicitly that this conclusion
remains  valid for the different parameter values
that were considered in tables \ref{table:power:f}
and \ref{table:power:varpars}.

\subsection{Comparison with results in the literature}
Refs.~\cite{Kashti:2008bw,Cuoco:2005yd,Kalashev:2007ph} explicitly give the
statistical power to rule out an isotropic CR distribution when
CR sources follow structure for other statistical tests,
thus allowing for a comparison with
our estimates.
In table \ref{table:powerlit} we compare the statistical
power of the  $Y$-test proposed here to the results of
Refs.~\cite{Kashti:2008bw,Cuoco:2005yd,Kalashev:2007ph}.
In obtaining our estimates we vary event numbers, threshold
energies, statistical significance, and other model parameters 
to match the values that were used in the quoted references.
(We however stick to the KKKST catalog for the distribution of galaxies).
Results between the different references should therefore not be
compared.
The results in the table show that the power of the $Y$-test,
as determined in this work,
is comparable to or larger than the power of the tests proposed
in the quoted references.

\TABLE{
\begin{tabular}{| c c c | c | c c | l |}
\hline
$E$ (EeV) & $N_{\rm ev}$ &  Acceptance & $\alpha $ & $P_Y$ & $P_{\rm ref}$ &  Reference \\
\hline
50 & 100 & PAO & 0.01 &  0.45 & 0.12 &  Ref.~\cite{Cuoco:2005yd}  \\
50 &  50 & PAO & 0.01 &  0.22 & 0.08 &  \\
\hline
40 & 100 & PAO & 0.05 & 0.46 & 0.26 &  Ref.~\cite{Kashti:2008bw} \\
60 & 31 & PAO & 0.05 & 0.56 & 0.25  &  ($X_C$; $\bar{s}_0 = 10^{-2}$ Mpc$^{-3}$ case)   \\
80 & 10 & PAO & 0.05 & 0.53  & 0.19  & \\ 
\hline
70 & 30 & Uniform & 0.05 & 0.93 & $\sim$0.95 & Ref.~\cite{Kalashev:2007ph} \\
\hline
\end{tabular}
\caption{\label{table:powerlit} Comparison of the statistical power of
the $Y$-test proposed in this work with  
results in the literature. The columns denote 
threshold energy, number of events,
assumed detector exposure, statistical significance,
power estimate in this work ($P_Y$) and in the literature ($P_{\rm ref}$), and a reference to the literature. 
In producing this table we have varied model parameters 
to match the values used in the quoted references.
Note that these power estimates depend both on the statistical test
and on the modeling  of CR arrival directions
(including the assumed galaxy catalog)
in the various studies.
}}

The power estimates in table \ref{table:powerlit} depend both
on the test itself and on the modeling of CR arrival directions.
If one is interested in a comparison of the 
tests \emph{per se}, the tests should be applied to exactly the same
event sets. We have done this for the $X_C$-statistic
put forward in Ref.~\cite{Kashti:2008bw} (denoted as $X$ in this paper)
in the previous subsections;
table \ref{table:power:DYXF}
shows how this test statistic compares to
the $Y$-test in performance.\footnote{We note that the power of the $X$-test
for 100 events above 40 EeV 
as reported in table \ref{table:power:DYXF} ($P_X= 0.79$)
is significantly larger
than the results given in Ref.~\cite{Kashti:2008bw} ($P_X = 0.26$).
This is primarily due to the different catalogs that were used;
we will come back to this in  Sect.~\ref{sec:discussion:X}.}

\section{Anisotropy in AGASA and PAO data}
\label{sec:pval}
In this section we assess the compatibility
of the publicly available data from AGASA
\cite{Takeda:1999sg, Hayashida:2000zr} and PAO  
\cite{Abraham:2007si} with an isotropic distribution
and predictions of the \model, with the local matter density in the Universe reconstructed
from the KKKST catalog.
The data sets are cut at $|b|>15\degree$, resulting in a total of 44 events
above 40 EeV for AGASA and a total of 19 events above 57 EeV for PAO.
The results presented here are primarily intended to illustrate the method
presented in this work,
the number of available events still being small
and the energy calibration uncertain.

\subsection{Method}
\TABLE{
\begin{tabular}{| c | c c c c | }
\hline
  & $D$ & $Y$ & $X$ & $F$  \\
\hline
\hline
value of test statistic in AGASA data & 0.087 & 0.050  & -0.59 & 0.19  \\
\hline
$p$-value for ``isotropy'' model & 0.86 & 0.61 & 0.88 & 0.20  \\ 
$p$-value for \modelsp & 0.98 & 0.20 &  0.018 &  0.45 \\ 
\hline
\end{tabular}
\caption{\label{table:pvals:AGASA} Values of the $D$-, $Y$-, $X$- and $F$-test statistics
for the 44 AGASA events with $|b|>15\degree$ and the corresponding $p$-values for
the ``isotropy'' model and the \modelsp
(for which we use the KKKST catalog
to reconstruct the local matter density).
These results are obtained using
$N_{\rm MC} = 10^5$  and $N_{\rm ref} = 10^5$ ($N_{\rm ref} = 10^6$ for $X$).}
}

\TABLE{
\begin{tabular}{| c | c c c c | }
\hline
  & $D$ & $Y$ & $X$ & $F$  \\
\hline
\hline
value of test statistic in PAO data & 0.25 & 0.18  & 6.3 & 0.16  \\
\hline
$p$-value for ``isotropy'' model & 0.15 & 0.20 & 0.035 & 0.83   \\ 
$p$-value for \modelsp & 0.58 & 0.66 &  0.85 & 0.88   \\ 
\hline
\end{tabular}
\caption{\label{table:pvals:PAO} Values of the $D$-, $Y$-, $X$- and $F$-test statistics
for the 19 PAO events with $|b|>15\degree$ and the corresponding $p$-values for
the ``isotropy'' model and the \modelsp  with the KKKST catalog. These results are obtained using
$N_{\rm MC} = 10^5$  and $N_{\rm ref} = 10^5$ ($N_{\rm ref} = 10^6$ for $X$).}
}

The $p$-values corresponding to the four statistical tests under consideration
are determined by integrating
the test statistic distributions in the model to be
tested over the range $R$ of possible
outcomes that are at least as extreme as the observed one
(see section \ref{sec:statistics}).
The test statistics are determined from eqs.
\eqref{eq:defts:D}$-$\eqref{eq:defts:X}, where
we compute the quantities $\mathcal{C}_{\rm ref} (\chi)$, 
$\mathcal{C}_{\rm ref} (l, b)$, 
and $N_{i, {\, \rm ref}}$ in the isotropy model and
the quantity $N_{i, {\, \rm alt}}$ in the \model.
(This is a matter of choice, and one could also consider
a test where the reference quantities are computed in the \model. We
do not pursue this possibility here.)
The numerical sampling of the test statistic distributions
is discussed in Sect.~\ref{sec:powercomparison:method}.

The exposure of both AGASA and PAO is essentially
geometrical at the highest energies. We model this exposure
following Ref.~\cite{Sommers:2000us},
adopting a detector latitude $a_0=36\degree$ ($-35\degree$)
and a maximum viewing angle $\theta_{m} = 45\degree$
($60\degree $) for AGASA (PAO).

\subsection{Results}
In table \ref{table:pvals:AGASA} we show the
values of the $D$-, $Y$-, $X$- and $F$-test statistics
for the AGASA data \cite{Takeda:1999sg, Hayashida:2000zr}, together with
the associated $p$-values for the ``isotropy'' model and the
\model. The $p$-values are all above 0.01 and
are therefore not conclusive.
The general trend is that the data are
more compatible with isotropy than with the \model. In particular,
the $X$-test statistic yields a $p$-value of $0.018$ for the AGASA
data and the \model, indicating mild incompatibility.
These results are compatible with 
the absence of large-scale anisotropies reported in Ref.~\cite{Takeda:1999sg}.

Table \ref{table:pvals:PAO} shows the values of the test statistics
for the PAO data \cite{Abraham:2007si} together with the $p$-values for the ``isotropy'' and
\model. Also here the $p$-values are larger than 0.01 and hence
inconclusive. 
All tests under consideration are fully compatible with
the \model. The $X$-test shows a very mild ($p=0.035$)
incompatibility between the
PAO data and the isotropy model predictions, while the other tests
are well compatible with isotropy.

It is not very surprising that an isotropic distribution
cannot be ruled out with
a strong significance with 19 events above 57 EeV when the
\modelsp is true: 
we have estimated in the previous section that
the probability to find a $p$-value smaller than
0.01 for the PAO data and the isotropy model
ranges from 3.3\% ($F$-test) to 26\% ($X$-test) in this case
(see table \ref{table:power:DYXF}).

We expect that the tests proposed in this work will 
constrain models of CR origin when more data become
available. We estimate that $\sim$$50$ events above 57 EeV
are required to have a  50\% probability of
excluding an isotropic distribution with significance 0.01
if the \modelsp is correct. (We obtained this results with the
procedure discussed in the previous section).
The required data can be accumulated by PAO and the Telescope Array (TA)
experiment \cite{Tokuno:2008zz}
within a few years. A determination of the optimal
search strategy (including optimization of the energy threshold)
requires an extensive study of the parameter space and is beyond
the scope of this paper.

\FIGURE{

\vspace{0.5cm}

\includegraphics[height=4.5cm]{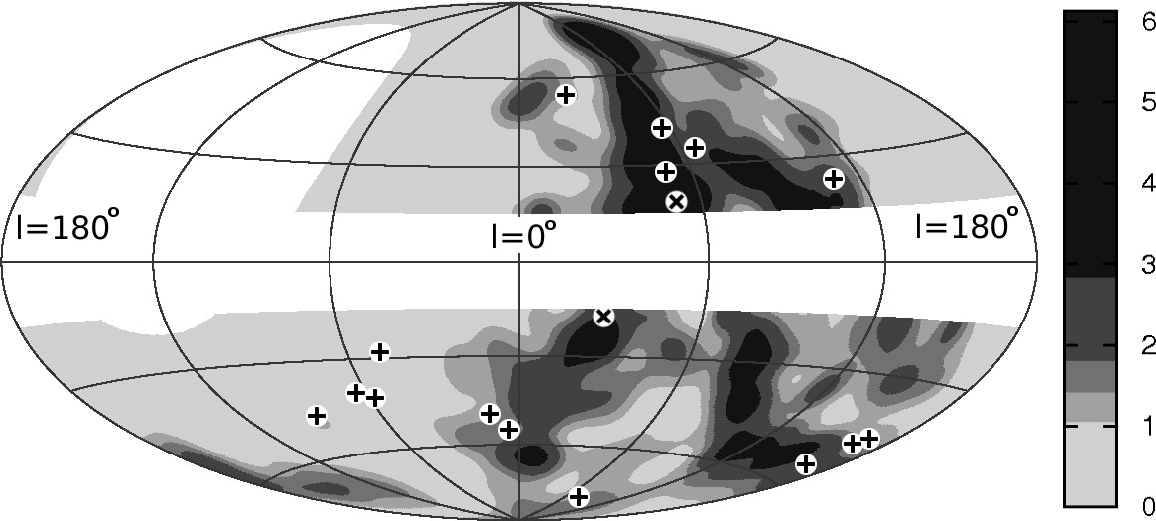}

\includegraphics[height=4.5cm]{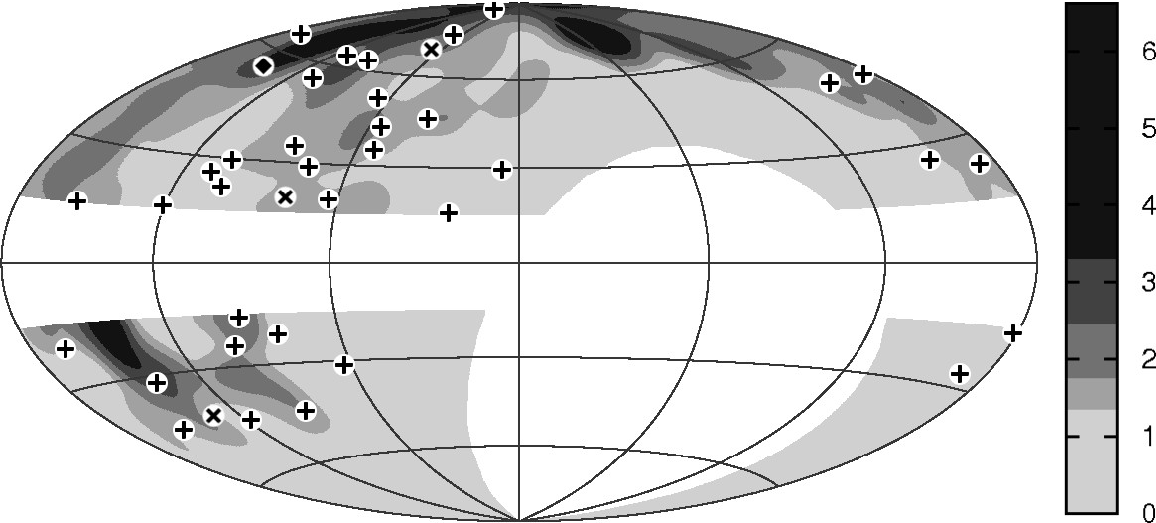}

\caption{\label{fig:PAacc} Aitoff projection of the sky 
(same coordinates as in Fig.~\ref{fig:skymap:galaxies})
showing the relative integral CR flux, modulated with the
relative detector exposure,
in the \model. 
 In modeling the CR flux we use the KKKST catalog 
to reconstruct the local matter density in the Universe.
The grayscale indicates the relative flux, such that
each band integrates to a fifth of the total flux.
The white area indicates the region outside the detector
field of view.
The `+', `$\times$', and `$\blacklozenge$' symbols show the arrival
directions of UHECRs, where `$\times$' indicates a doublet
and  `$\blacklozenge$' a triplet. Upper panel: PAO model flux
and data above 57 EeV; bottom panel: AGASA model
flux and data above 40 EeV. Data are from
Refs. \cite{Abraham:2007si, Takeda:1999sg, Hayashida:2000zr}.}
}

\section{Discussion}
\label{sec:discussion}
In this section we discuss some aspects of
the tests proposed in this work, and 
we compare the results
presented in sections \ref{sec:powercomparison} and \ref{sec:pval}
to estimates presented in Ref.~\cite{Kashti:2008bw}.

\subsection{Blindness}
We have shown in Sect.~\ref{sec:powercomparison}
that the $D$- and $Y$-tests are very
efficient in discriminating between the \modelsp and an
isotropic distribution of UHECRs, especially for the highest energy part
of the UHECR spectrum where the number of events is small and the flux
map exhibits more contrast. The sensitivity comes at a price: one may
imagine deviations from the \modelsp to which our test is
blind. The existence of such deviations is clear from
Fig.~\ref{fig:skymap:example} (which corresponds to the example discussed in the
introduction): the histogram on the left does not
change if one moves the events leaving them on the same flux contour
(i.e. the same grayscale). 
More generally, the $D$- and $Y$-test statistics as defined in eqs. 
\eqref{eq:defts:D} and \eqref{eq:defts:Y} are unchanged if we move 
CR events along contours of equal $\chi$.
In particular, one may group all events that fall on the same flux contour
in one place without changing the $D$- and $Y$-test statistics.
Although this would certainly increase the anisotropy, it would
not be noticed by these tests.
The $X$-test proposed in Ref.~\cite{Kashti:2008bw}
suffers from the same kind of blindness, as it is invariant under
redistributions of events over bins with equal fraction
$N_{i, {\, \rm ref}}/N_{i, {\, \rm alt}}$.

\subsection{Comparison with earlier results for the $X$-test}
\label{sec:discussion:X}
Here we compare our results for the statistical power and  $p$-values of 
the $X$-test with results
presented by Kashti and Waxman in Ref. 
\cite{Kashti:2008bw}, where this statistic was
introduced.\footnote{It is denoted as $X_C$ in Ref.~\cite{Kashti:2008bw}.}
We compare our results with the largest source density
considered in Ref.~\cite{Kashti:2008bw}, because we are interested in
the many-source regime, and with the ``unbiased'' model because this resembles
our setup most closely.

Our estimates for the 
statistical power to reject an isotropic distribution when the
\modelsp is true are significantly larger
than the estimates in Ref. \cite{Kashti:2008bw}, which is to say
that we find a much better discrimination between the two models.
For scenario I ($N_{\rm ev} = 100$ events above $E=40$ EeV) we find a power 
$P_X = 0.79$ (significance 5\%; see table \ref{table:power:DYXF}),
whereas Kashti and Waxman find $P_X = 0.26$ (same significance;
their table
II).
We have verified that our estimates are also larger for other
values of $N_{\rm ev}$ and $E$ for which power estimates
are given in Ref.~\cite{Kashti:2008bw};
these do not include our scenarios II and III though.
Although our model assumptions differ in more than one respect,
the observed difference appears to be
primarily due to the different catalogs that were 
used in modeling  CR arrival directions, namely the PSCz
catalog  \cite{Saunders:2000af} in Ref.~\cite{Kashti:2008bw} and the KKKST catalog in this work
(Sect.~\ref{sec:srcdist}).
To show this we have recomputed the statistical
power for the $X$-test using the PSCz catalog and find that
the power reduces to $P_X = 0.51$. When we also change model parameters
and cuts to match the values used in Ref.~\cite{Kashti:2008bw} (i.e., 
$p=2.0$, strong source evolution, cut the galactic plane
above $|b|=12\degree$) we find $P_X = 0.39$, in reasonable
agreement with the result $P_X = 0.26$ obtained in Ref.~\cite{Kashti:2008bw}.
(Comparing our results to data files referred to 
in Ref.~\cite{Kashti:2008bw}, we trace the fact that
our estimates are somewhat higher to
a stronger contrast in the matter densities derived from the PSCz catalog).
The relatively strong impact of the source catalog on 
the statistical power is in keeping with
results obtained for the $Y$-test in section 
\ref{sec:powercomparison} (see table
\ref{table:power:varpars}).

Now we compare the $p$-values for an isotropic model
distribution and the PAO data
between this study and Ref.~\cite{Kashti:2008bw}.
Our $p$-value of $0.035$ (see table \ref{table:pvals:PAO})
is almost an order of magnitude larger than
the value of $0.004$ reported in Ref.~\cite{Kashti:2008bw}
(their table III), meaning
that we find a much better agreement between the
data and an isotropic CR distribution.
At first sight this may appear to be in contradiction with the fact that our
power  estimates are larger than those in Ref.~\cite{Kashti:2008bw}:
the larger statistical power leads us to expect
--- if the \modelsp 
describes nature correctly ---
that we would find a stronger incompatibility
between ``isotropy'' and PAO data
(more precisely, we have better chances of finding a strong
incompatibility). 
This discrepancy can be explained if
the true CR source density contrasts are larger than the density
contrasts in the PSCz catalog used in Ref.~\cite{Kashti:2008bw},
so that the model predictions based on that catalog are not accurate.
This may (partly) be due to selection effects, as it has been pointed out that
the limited angular resolution
of the IRAS instrument used to compile the catalog may lead
to a systematic underestimate of the number density of galaxies
in high-density regions (see discussion in Ref.~\cite{Kalashev:2007ph}).
It may also indicate that CR sources are intrinsically
biased with respect to the infrared galaxy
sample underlying the PSCz catalog. 
The results presented in  Ref.~\cite{Kashti:2008bw} lend
further support to the hypothesis that
CR source density contrasts are larger than those derived
from the PSCz catalog:
The $p$-value
for the PAO data and the ``unbiased'' model (in which CR sources trace the distribution
of matter) in Ref.~\cite{Kashti:2008bw} is also small, viz. $p=0.021$ (table III in their work),
which indicates a mild discrepancy between the PAO data and the
model predictions based on the PSCz catalog. 
(We find a $p$-value of 0.85, and hence perfect agreement, 
when we apply the $X$-test to the PAO data and the
\modelsp using the KKKST catalog;
see table \ref{table:pvals:PAO}).
In fact, the PAO value
of the $X$-test statistic in Ref.~\cite{Kashti:2008bw} is larger than
the average value predicted in their ``unbiased'' model, which in turn
is larger than the average value predicted under the assumption of
an isotropic event distribution. This suggests that the data exhibit more
contrast than the ``unbiased'' model predictions. 
From the same observation,
the authors of Ref.~\cite{Kashti:2008bw}
conclude that the PAO data show a preference towards
their ``biased'' model,
which has more contrast than the ``unbiased'' model.
Future CR data will  provide further clues as to
the actual distribution of sources and the relative strength
of over- and underdensities.

\section{Conclusion}
\label{sec:conclusion}
In this paper we have presented a new and powerful method to test
models of cosmic-ray origin that predict a continuous distribution of CRs
over the sky.
A prominent model of this kind is the ``\model'',
a simple yet general model that predicts that local matter
overdensities are reflected
in the distribution of CRs.
Its main assumptions,
a continuous distribution of sources
tracing the distribution of matter in the Universe
and small CR deflections,
represent the limiting case of many detailed
scenarios. We therefore consider this model
as a useful benchmark model in CR anisotropy studies.
The local matter density in the Universe, which is required as input for
the \model, is derived from a galaxy catalog.
For this purpose we primarily use
the catalog 
compiled by Kalashev et al.
\cite{Kalashev:2007ph}, a complete sample of galaxies up to 270 Mpc (the KKKST catalog; see Sect.~\ref{sec:srcdist}). We have
however recomputed some of our results using the PSCz catalog \cite{Saunders:2000af}, which was
used in various earlier studies (e.g., Ref.~\cite{Kashti:2008bw}),
for comparison.

We analyzed in detail (Sect.~\ref{sec:powercomparison}) the statistical power of the two
proposed tests (denoted as the $D$- and $Y$-tests)
to reject an isotropic CR distribution when
the \modelsp is true (with matter densities reconstructed from the KKKST catalog).
We find that the proposed tests are competitive or superior
in power
to other tests in the literature. They are especially
sensitive in the physically interesting
case of  relatively few events with very high energies, viz.
$\mathcal{O} (10)$ events above $E=100$ EeV
(the test statistic proposed in Ref.~\cite{Kashti:2008bw} provides
a more powerful discrimination in the case of many events
with lower threshold energy, $E \gtrsim 40$ EeV).
The proposed tests are completely binless, 
which is an important advantage because binning introduces an ambiguity that
complicates the statistical analysis.

We have applied our tests to the AGASA data
above 40 EeV \cite{Takeda:1999sg, Hayashida:2000zr}
and the PAO data above 57 EeV \cite{Abraham:2007si}, and
found that the present
data are compatible with both the \modelsp (with matter densities reconstructed from the KKKST catalog)
and with an
isotropic CR distribution (Sect.~\ref{sec:pval}). The 
general trend is that the AGASA
data show a greater compatibility with an isotropic
distribution, whereas the PAO data favor
the \model. We expect that
the proposed tests will provide meaningful constraints on models when more
data become available. 
We estimate that $\sim$$50$ events above 57 EeV
are required to have a  50\% probability of
excluding an isotropic distribution with significance 0.01
if the \modelsp is correct.
The required data can be accumulated by PAO and TA
within a few years. 

Summarizing our method, both the $D$- and $Y$-
tests are based on the (continuous)
distribution of a parameter $\chi$ 
--- which  is a measure of the amount of matter
in the nearby Universe  ---
over a set of observed CRs
(Sect.~\ref{sec:chi}). 
In the \modelsp
the probability that a CR comes from an overdense region
is higher than  from an underdense region, so that
the $\chi$-parameter takes preferentially high values.
For an isotropic distribution there is no such correlation and
$\chi$ is essentially a random variable.
Correlating the distribution of $\chi$ in a set of observed
CR events with model distributions then gives a measure of
the (in)compatibility between observations and model predictions.
This last step may be done with the standard Kolmogorov-Smirnov
test ($D$-test) or a modified version thereof proposed in this
work ($Y$-test; Sect.~\ref{sec:TS}). The latter is shown to be
somewhat more powerful in discriminating between the \modelsp 
and an isotropic CR distribution.

 Reconstructing the matter density
in the Universe from a galaxy catalog, as done in this study, necessarily introduces
some ambiguity and uncertainties. 
For the KKKST catalog, which we have primarily used, the dominant source of errors
is in the photometric redshift estimates for galaxies beyond 30 Mpc.
The associated uncertainties are acceptable given the other uncertainties
we are faced with, most notably in the energy resolution of
present CR experiments and in modeling CR energy loss in the Universe.
For the purpose of this study, the most important difference between the
KKKST catalog and the PSCz catalog  is that the
former exhibits stronger density contrasts, which improves the prospects of
discriminating between the
\modelsp and an isotropic CR distribution.
Whether or not actual CR sources exhibit such a strong density contrast 
remains an open question, to be answered by observations.
We would like to point out that the increased contrasts cannot be an
artefact of the random errors associated with photometric redshift estimates.
Such errors will, on average, bring
sources closer. Keeping the CR horizon fixed, this results in
a more homogeneous flux distribution rather than a more contrasted one (as may be easily seen by noting the
equivalence with increasing the horizon while keeping distances fixed).

Current and future data can test model predictions 
based on a given galaxy catalog. 
Using the present data from PAO, we argued
in Sect.~\ref{sec:discussion:X} that the CR source density seems to
exhibit stronger contrasts
than what is derived from the PSCz catalog, while the data
are fully compatible with the KKKST catalog. This picture is however not supported
by the AGASA data, which favor an isotropic model.
In the future it may become possible to investigate the spatial distribution of CR
sources in a greater amount of detail
by comparing CR data with predictions from catalogs based on
various galaxy types with different clustering properties.
This would require more, and more precise, CR data as well as improved galaxy catalogs.

We comment on some possible improvements to and extensions 
of the proposed tests. A first possibility is to include the energies
of individual events  rather than a single threshold energy in the analysis.
A preliminary investigation has shown that this can improve the statistical
power, but more work is needed to study this idea in detail. Second,
we note that it is possible to choose the statistical parameter
$\chi$ in a different manner than eq. \eqref{eq:def:chi}, e.g.
to be equal to the detector exposure (so that $\chi$ will take
preferentially high values in an isotropic model). We have not
investigated this possibility
here. Finally the estimates for proton energy-loss lengths  may
be updated using more sophisticated numerical propagation routines
when future data call for a higher degree of precision.

\acknowledgments
We thank Sergey Troitsky for valuable discussions and for 
providing us with the KKKST catalog, and Alessandro Cuoco,
Michael Kachelrie\ss{}, Tamar Kashti, and Eli Waxman for comments.
H.K. would like to thank Thomas Lessinnes for illuminating discussions on 
hypothesis testing.
H.K. and P.T. are supported by Belgian Science Policy under IUAP VI/11
and by IISN. The work of P.T. is supported in part by the
FNRS, contract 1.5.335.08.

\bibliographystyle{JHEP}
\bibliography{refs}

\end{document}